\documentclass[11pt,a4paper]{article}
\usepackage{amsmath}
\usepackage{amsfonts}
\usepackage{color}
\usepackage{graphicx}
\graphicspath{{figures/}}
\usepackage{dcolumn}
\usepackage{bm}
\usepackage[
bookmarks,bookmarksnumbered,colorlinks=true,anchorcolor=blue,
linkcolor=blue,urlcolor=blue,citecolor=blue,
breaklinks=true]{hyperref}
\usepackage{geometry}
\usepackage{hep}
\usepackage{extarrows}
\usepackage{authblk}
\usepackage{amssymb}
\usepackage[numbers,sort&compress]{natbib}
\usepackage{cite}
\usepackage{pifont}
\usepackage{tikz}
\usepackage{sansmath}
\usepackage{mathrsfs}
\usetikzlibrary{shadings,intersections,snakes}
\usepackage{verbatim}
\numberwithin{equation}{section}
\usepackage{tikz}
\usetikzlibrary{calc}

\date{\today}

\begin{document}

\title{\bf Holographic Schwinger-Keldysh field theory of SU(2) diffusion}

\author[]{Yanyan Bu \thanks{yybu@hit.edu.cn}~}

\author[]{Xiyang Sun \thanks{xysun@stu.hit.edu.cn (correspondence author)}~}

\author[]{Biye Zhang \thanks{zhangbiye@hit.edu.cn}}

\affil[]{\it School of Physics, Harbin Institute of Technology, Harbin 150001, China}

\maketitle

\begin{abstract}

We construct effective field theory for SU(2) isospin charge diffusion, based on holographic Schwinger-Keldysh contour \cite{Glorioso:2018mmw}. The holographic model consists of a probe SU(2) gauge field in a doubled Schwarzschild-AdS$_5$ geometry. Accurate to first order in derivative expansion, we analytically compute the effective action up to quartic order in dynamical variables. The effective theory contains both non-Gaussianity for noises and nonlinear interactions between noises and dynamical variables. Moreover, the effective theory captures both thermal and quantum fluctuations, which perfectly satisfy dynamical Kubo-Martin-Schwinger (KMS) symmetry at quantum level. Interestingly, the dynamical KMS symmetry, which is crucial in formulating non-equilibrium effective field theory for a quantum many-body system, is found to have a nice holographic interpretation.

\end{abstract}

\newpage

\tableofcontents

\allowdisplaybreaks

\flushbottom

\section{Introduction}

Hydrodynamics is believed to be an effective description of a quantum many-body system in the long distance long time regime. Classical hydrodynamics is formulated as a set of deterministic partial differential equations (PDEs), which stands for conservation laws of stress-energy tensor and internal currents. However, classical hydrodynamics does not capture effects of statistical and/or quantum fluctuations. In stochastic formulation of hydrodynamics, fluctuations are modelled by random dissipative fluxes (with Gaussian distributions), added to the stress-energy tensor and internal currents. Resultantly, the deterministic conservation laws become stochastic Langevin-type PDEs.

Recently, hydrodynamics has been reformulated as a non-equilibrium effective field theory (EFT) \cite{Crossley:2015evo,Glorioso:2016gsa,Glorioso:2017fpd,Haehl:2015foa,Haehl:2015uoc,Haehl:2018lcu}\footnote{ See \cite{Endlich:2012vt,Kovtun:2014hpa,Nicolis:2013lma,Harder:2015nxa,
Grozdanov:2013dba} for early attempts on this subject.} (see \cite{Liu:2018kfw} for a pedagogical review) by utilizing Schwinger-Keldysh (SK) formalism, entirely based on symmetry principles. In contrast to stochastic formulation, hydrodynamic EFT incorporates dissipations and fluctuations in a systematic manner. Particularly, fluctuation-dissipation theorem (FDT) is implemented as a dynamical KMS symmetry \cite{Glorioso:2016gsa,Glorioso:2017fpd} satisfied by the hydrodynamic effective action. The dynamical KMS symmetry acts on dynamical variables of the hydrodynamic EFT. Thus, hydrodynamic EFT provides an ideal framework for investigating fluctuation effects in a variety of physical problems, see e.g. \cite{Chen-Lin:2018kfl,Chao:2020kcf,Landry:2020ire,Jain:2020zhu,Jain:2020hcu,Sogabe:2021wqk,Sogabe:2021svv} for recent progress.

The development of hydrodynamic EFT has been largely inspired by the AdS-CFT correspondence \cite{Maldacena:1997re,Gubser:1998bc,Witten:1998qj}, which conjectures that strongly coupled large $N_c$ (the number of colors) gauge theory is equivalent to a weakly coupled gravitational theory in higher dimensional asymptotic AdS space. Essentially, deriving hydrodynamic EFT from gravity amounts to implementing holographic Wilsonian renormalization group \cite{Crossley:2015tka,deBoer:2015ija,Heemskerk:2010hk,Faulkner:2010jy,Sin:2011yh} in dual gravity. A first successful derivation was presented for a U(1) charge diffusion \cite{deBoer:2018qqm,Glorioso:2018mmw}. While \cite{deBoer:2018qqm} adopted real-time formalism for AdS-CFT correspondence \cite{Herzog:2002pc,Skenderis:2008dh,Skenderis:2008dg}\footnote{This real-time prescription has been used to compute higher-point correlation functions \cite{Barnes:2010jp,Arnold:2010ir,Arnold:2011ja,Arnold:2011hp}.}, the work \cite{Glorioso:2018mmw} achieved the goal by proposing a holographic prescription for Schwinger-Keldysh (SK) closed time contour. The results of \cite{deBoer:2018qqm,Glorioso:2018mmw} are in perfect agreement with that constructed within the non-equilibrium EFT framework \cite{Crossley:2015evo}. Recently, the holographic SK contour \cite{Glorioso:2018mmw} attracted a lot of attention in various holographic settings \cite{Chakrabarty:2019aeu,Jana:2020vyx,Chakrabarty:2020ohe,Loganayagam:2020eue,
Loganayagam:2020iol,Ghosh:2020lel,Bu:2020jfo,Bu:2021clf,He:2021jna,Bu:2021jlp}.

In this work we extend studies of \cite{Glorioso:2018mmw,Bu:2020jfo} and present a holographic derivation of hydrodynamic EFT for a SU(2) isospin charge diffusion\footnote{Classical hydrodynamics with an internal SU(2) symmetry has been considered, for instance, in \cite{Torabian:2009qk,Eling:2010hu,Neiman:2010zi,Hoyos:2014nua,Fernandez-Melgarejo:2016xiv,Dantas:2019rgp}. }, using the holographic SK contour \cite{Glorioso:2018mmw}. This extension is highly nontrivial, mainly due to nonlinearity of the theory. Indeed, for a linear bulk theory, in order to consistently cover both ingoing mode (dual to dissipation) and outgoing mode (dual to fluctuation), the bulk fields inevitably exhibit logarithmic divergences near the event horizon \cite{Glorioso:2018mmw,Bu:2020jfo}. When nonlinearities are present, this logarithmic behavior near the horizon will bring in technical complications in the derivation of boundary action, which is one of the main problems we will address in this study. Apart of this, there is a subtle issue \cite{Glorioso:2018mmw} regarding the non-commutativity between hydrodynamic limit and near horizon limit. In this work, we would like to further explore this point and develop a more solid way to extract hydrodynamic EFT action on the boundary.

Our study is partly motivated by a recent publication \cite{Glorioso:2020loc}, which developed an EFT for hydrodynamics with global non-Abelian symmetries. Ref.~\cite{Glorioso:2020loc} assumes two main approximations: truncating the effective action to quadratic order in noise variables, and imposing dynamical KMS symmetry in the classical statistical limit (i.e., $\hbar \to 0$)\footnote{Implementation of dynamical KMS symmetry at quantum level has been considered for U(1) charge diffusion \cite{Jensen:2017kzi,Jensen:2018hhx}  and in formulating an EFT for maximally quantum chaotic system \cite{Blake:2017ris,Blake:2021wqj,Abbasi:2021fcz}.}. While the former approximation does not completely kill nonlinear interactions between noises and dynamical variables, it does ignore non-Gaussianity in noise. Recently, Jain and Kovtun clarified importance of non-Gaussian noises in quantifying (non-)universality of hydrodynamics \cite{Jain:2020zhu}: non-Gaussian noise does have non-negligible signals in hydrodynamic correlation functions.

The second approximation made in \cite{Glorioso:2020loc} corresponds to ignoring quantum fluctuations. While this works well from hydrodynamic EFT perspective \cite{Crossley:2015evo}, it becomes unsatisfactory on holographic side: a clear splitting between quantum fluctuations and statistical fluctuations would be impossible for a holographic field theory \cite{deBoer:2018qqm}. More precisely, for a holographic theory, the mean free path is $\sim \hbar /T$, which implies that derivative expansion would unavoidably bear quantum fluctuations \cite{deBoer:2018qqm}. In other words, the large $N_c$ limit of holography does not necessarily mean classicalization of the boundary system, which is another conceptual issue we would like to clarify through present study. Hopefully, elaboration on this matter will shed light on the imposition of dynamical KMS symmetry at quantum level for the construction of hydrodynamic EFT for a dissipative fluid, which is still an open question \cite{Glorioso:2016gsa,Glorioso:2017fpd}. In addition, we are interested in understanding to what the dynamical KMS symmetry proposal \cite{Glorioso:2016gsa,Glorioso:2017fpd} corresponds in the dual gravity theory.

Therefore, via holographic technique, we aim at deriving a more comprehensive EFT for SU(2) isospin charge diffusion, in which the two approximations undertaken in \cite{Glorioso:2020loc} will be relaxed. Phenomenologically, the present study will be of potential relevance to the study of QCD plasma in the chiral limit \cite{Torabian:2009qk,Grossi:2020ezz}: on the one hand, the global SU(2) symmetry can mimic flavor symmetry of QCD in two-flavor approximation in a late stage of RHIC plasma; on the other hand, this study would be insightful in understanding fluctuations in pion fluid at finite temperature.

Before diving into detailed calculations, we summarize our main results:

\noindent $\bullet$ Limited to first order in derivative expansion, we derive the hydrodynamic effective action for SU(2) diffusion, valid to quartic order in hydrodynamic fields. We analytically compute all the parameters in the effective action. See equations \eqref{Leff_quadratic}, \eqref{Leff_cubic}, \eqref{Leff_quartic_LO}, and \eqref{Leff_quartic_NLO} (or the more compact ones \eqref{Leff_quadratic_cov}-\eqref{Leff_quartic_cov}) for detailed results.

\noindent $\bullet$ We reveal that our effective action satisfies dynamical KMS symmetry at the quantum level, say, both thermal and quantum fluctuations are accounted for in the effective action, see subsection \ref{holo_KMS_symmetry}.

\noindent $\bullet$ We present a holographic interpretation for dynamical KMS symmetry proposed in \cite{Glorioso:2016gsa,Glorioso:2017fpd}: under bulk KMS transformation, the bulk fields living on the lower (upper) branch of the holographic contour will transform in an analogous way as those living on the forward (backward) branch of the SK contour in boundary theory. See subsection \ref{holo_KMS_symmetry} for more discussions.

The rest of this paper will be organized as follows. In section \ref{holo_setup} we present the holographic setup: a probe SU(2) gauge field in doubled Schwarzschild-AdS$_5$ geometry. In addition, we outline basic strategy of deriving boundary effective action from bulk dynamics. Section \ref{bulk_solution} is the main part of this work. First, we solve classical bulk dynamics in the partially on-shell sense; then, we compute the partially on-shell bulk action by implementing radial contour integrals, which gives the boundary effective action; finally, we check dynamical KMS symmetry of our effective action, and explore a holographic interpretation for it. In section \ref{summary}, we make a summary and discussions. Appendices \ref{YM_eom_explicit} and \ref{generic_structure_solution} supplement further calculational details.

\section{Holographic setup} \label{holo_setup}

{\bf Notation convention}: throughout this paper, we use the upper-case Latin letters ``$M,N,\cdots$'' to denote bulk spacetime indices, the Greek letters ``$\mu, \nu, \cdots$'' for boundary indices, the Italic lower-case Latin letters ``$i,j,\cdots$'' for spatial coordinates, and the upright lower-case Latin letters ``$\rm a,b, \cdots$'' for SU(2) flavor indices.

We consider a non-Abelian SU(2) gauge theory in Schwarzschild-AdS$_5$ geometry, whose dynamics is described by Yang-Mils action:
\begin{align}
S_0 = - \frac{1}{2} \int d^5x \sqrt{-g}\, {\rm Tr}(F^2) = - \frac{1}{4} \int d^5 x \sqrt{-g}\, F_{MN}^{\rm a} F^{{\rm a} MN}, \label{S0}
\end{align}
where Yang-Mills field and its field strength are matrix-valued
\begin{align}
& C= C_M dx^M = C_M^{\rm a} t^{\rm a} dx^M, \quad  F =dC = F_{MN} dx^M \wedge dx^N = F_{MN}^{\rm a} t^{\rm a} dx^M \wedge dx^N, \nonumber \\
& F_{MN}^{\rm a} = \nabla_M C_N^{\rm a} - \nabla_N C_M^{\rm a} + \epsilon^{\rm abc} C_M^{\rm b} C_N^{\rm c}, \qquad t^{\rm a} = \frac{1}{2} \tau^{\rm a},
\end{align}
where $\tau^{\rm a}$'s are the Pauli matrices, and repeated indices mean summation. Here, we take the convention for normalization ${\rm Tr}(t^{\rm a} t^{\rm b}) = \frac{1}{2}\delta^{\rm ab}$. In ingoing Eddington-Finkelstein (EF) coordinate system $x^M= (r,v, x^i)$, the metric of Schwarzschild-AdS$_5$ geometry is given by
\begin{align}
ds^2 = g_{MN} dx^M dx^N = 2 dvdr - r^2f(r) dv^2 + r^2 \delta_{ij} dx^idx^j, \qquad i,j=1,2,3, \label{Schw-AdS5}
\end{align}
where $f(r)= 1-r_h^4/r^4$ with $r_h$ the horizon radius (the AdS radius is set to unity). The Schwarzschild-AdS$_5$ has Hawking temperature $T= r_h/\pi$, which is identified as the temperature for boundary theory. In asymptotic AdS space, we also need a counter-term action \cite{Round:2010kj}
\begin{align}
S_{\rm ct} = \frac{1}{4}\log r \, \int d^4x \sqrt{-\gamma} F_{\mu\nu}^{\rm a} F^{{\rm a} \mu \nu} \bigg|_{r= \infty}, \label{Sct}
\end{align}
which is written down based on minimal subtraction scheme. Here, $\gamma$ is the determinant of induced metric $\gamma_{\mu\nu}$ on the boundary $r=\infty$.

\begin{figure}[htbp]
\centering
\includegraphics[width=0.8\textwidth]{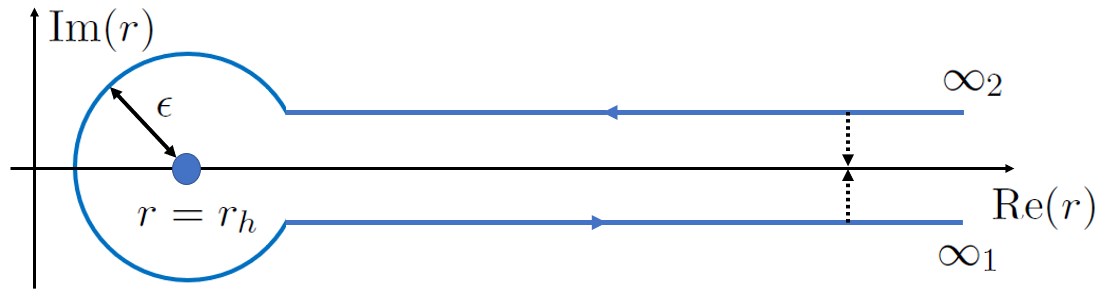}
\caption{Gravity dual of the SK time contour \cite{Glorioso:2018mmw}: complexified radial coordinate and analytical continuation around the horizon $r_h$. }
\label{holographic_SK_contour}
\end{figure}

For the purpose of incorporating fluctuation and dissipation in an action principle, the boundary system shall be placed on the SK time contour \cite{Kamenev2011}. A gravity dual of the SK time contour is proposed in \cite{Glorioso:2018mmw}, which complexifies the radial coordinate $r$ of \eqref{Schw-AdS5} and analytically continues it around the event horizon $r=r_h$, see Figure \ref{holographic_SK_contour}.

Here, we explain the basic strategy of deriving hydrodynamic effective action from AdS gravity. This has been originally formulated in \cite{Crossley:2015tka} for the problem of stress tensor (see also \cite{Glorioso:2018mmw,Bu:2020jfo} for U(1) charge diffusion problem), based on early attempts \cite{Heemskerk:2010hk,Faulkner:2010jy,Nickel:2010pr}. The starting point is the holographic dictionary \cite{Gubser:1998bc,Witten:1998qj}:
\begin{align}
Z_{\rm CFT} = Z_{\rm AdS}.
\end{align}
The CFT partition function $Z_{\rm CFT}$ may be presented as a path integral over gapless modes (collectively denoted by $X$) in the low energy EFT:
\begin{align}
Z_{\rm CFT}= \int [DX] e^{i S_{eff}[X]}, \label{Z_CFT}
\end{align}
where $S_{eff}$ is the desired effective action.
The AdS partition function $Z_{\rm AdS}$ is a path integral over bulk fields\footnote{In the probe limit, it is valid to ignore dynamics of bulk spacetime.},
\begin{align}
Z_{\rm AdS}= \int [D C_M^\prime] e^{i S_0[C_M^\prime] + i S_{\rm ct}}, \label{Z_AdS}
\end{align}
which will be computed in the saddle point approximation, i.e., based on solving classical dynamics of the bulk Yang-Mills theory \eqref{S0}. In \eqref{Z_AdS} the primed configuration $C_M^\prime$ corresponds to no gauge-fixing for the bulk gauge theory.

Thus, in order to obtain $S_{eff}$ from bulk theory we shall identity gapless mode of low energy EFT, whose gravity dual shall not be integrated out in the gravity partition function \eqref{Z_AdS}. This can be achieved by carefully examining gauge symmetry in the bulk, as illustrated for U(1) case by Nickel and Son \cite{Nickel:2010pr} (see also \cite{Glorioso:2018mmw,Bu:2020jfo,Bu:2021clf}). Through a bulk gauge transformation
\begin{align}
C_M^\prime \to C_M = U(\Lambda) (C_M^\prime + i \nabla_M ) U^{\dagger}(\Lambda), \qquad U(\Lambda) = e^{i \Lambda^{\rm a} t^{\rm a}}, \label{bulk_gauge_trans}
\end{align}
a generic configuration $C_M^\prime$ can be brought into any gauge-fixed form $C_M$, for example, the one with radial component fixed as $C_r =0$. The bulk gauge transformation parameter $\Lambda^{\rm a}(x^M)$ is determined by solving gauge-fixing condition. At the AdS boundary, \eqref{bulk_gauge_trans} takes the form:
\begin{align}
A_\mu^\prime(x^\alpha) \to \mathcal U(\varphi) \left[ A_\mu^\prime(x^\alpha) + i \partial_\mu \right] \mathcal U^{\dagger}(\varphi), \qquad \mathcal U(\varphi) = e^{i \varphi^{\rm a} t^{\rm a}}, \qquad \varphi^{\rm a}(x^\mu) \equiv \Lambda^{\rm a}(r=\infty,x^\mu),
\end{align}
where $A_\mu^\prime$ is boundary value of $C_\mu^\prime$. Thus, the gravity partition function \eqref{Z_AdS} can be equivalently expressed as a path integral over $C_M$ (gauge-fixed configuration) and the gauge transformation parameter $\Lambda$:
\begin{align}
Z_{\rm AdS} & = \int [D \Lambda] [D C_\mu] e^{iS_0[C_M, \Lambda] + i S_{\rm ct}} \nonumber \\
& \simeq \int [D \varphi] [D C_\mu] e^{iS_0[C_\mu, \varphi] + i S_{\rm ct}}, \label{Z_AdS1}
\end{align}
where in the last equality we ignored an overall normalization constant given that this will not affect results of physical quantities. Once $C_\mu$ is integrated out, \eqref{Z_AdS1} is put into the desired form of \eqref{Z_CFT},
\begin{align}
Z_{\rm AdS} = \int [D \varphi] e^{iS_0[C_\mu[\varphi], \varphi]|_{\rm p.o.s} + i S_{\rm ct}},
\end{align}
so that the boundary effective action is read off from bulk action
\begin{align}
S_{eff} = S_0[C_\mu[\varphi], \varphi]|_{\rm p.o.s} + S_{\rm ct}. \label{Seff_formula}
\end{align}
Here, under the philosophy of \cite{Nickel:2010pr,Glorioso:2018mmw,Crossley:2015evo}, we identify $\varphi$ as the gapless mode of low energy EFT. Note $S_0[C_\mu[\varphi], \varphi]|_{\rm p.o.s}$ is partially on-shell bulk action to be obtained by plugging classical solution for $C_\mu$ into Yang-Mills action \eqref{S0}. It is important to stress that by partially on-shell, we determine profile for $C_\mu$ by solving dynamical equations of motion (EOMs) only. Particularly, we will leave aside the constraint equation. Such a partially on-shell formalism was ever invented in \cite{Bu:2014sia,Bu:2014ena,Bu:2015ika,Bu:2015bwa,Bu:2015ame} to resum (linear) all-order derivatives in off-shell hydrodynamic constitutive relations. Below we elaborate on this point.

Under the variation
\begin{align}
C_M^{\rm a} \to C_M^{\rm a} + \delta C_M^{\rm a} \Longrightarrow \delta F_{MN}^{\rm a} = \nabla_M \delta C_N^{\rm a} - \nabla_N \delta C_N^{\rm a} + \epsilon^{\rm abc} \delta C_M^{\rm b} C_N^{\rm c} + \epsilon^{\rm abc} C_M^{\rm b} \delta C_N^{\rm c},
\end{align}
the Yang-Mills action \eqref{S0} varies as
\begin{align}
\delta S_0 = - \int d^5x \sqrt{-g} \left[ \nabla_M \left(\delta C_N^{\rm a} F^{{\rm a} MN} \right) - \delta C_N^{\rm a} \nabla_M F^{{\rm a} MN} + \delta C_N^{\rm a} \epsilon^{\rm abc} C_M^{\rm c} F^{{\rm b}NM} \right].
\end{align}
Immediately, we read off the bulk EOMs:
\begin{align}
\nabla_M F^{{\rm a} MN} + \epsilon^{\rm abc} C_M^{\rm b} F^{{\rm c} MN}=0. \label{YM_eqn}
\end{align}
Throughout this work, we will take the gauge convention
\begin{align}
C_r^{\rm a} = - \frac{C_v^{\rm a}}{r^2f(r)},  \label{radial_gauge_Schw}
\end{align}
which means in Schwarzschild coordinate system the radial component of bulk gauge field is fixed to zero \cite{Bu:2020jfo,Bu:2021clf}.
The motivation for taking such a gauge choice is to realize time-reversal symmetry in a simple way, see \cite{Bu:2020jfo}.
Accordingly, the dynamical components of bulk EOMs are \cite{Bu:2020jfo,Crossley:2015tka}
\begin{align}
& \nabla_M F^{{\rm a} Mi} + \epsilon^{\rm abc} C_M^{\rm b} F^{{\rm c} Mi}=0, \nonumber \\
& \nabla_M F^{{\rm a} Mv} + \epsilon^{\rm abc} C_M^{\rm b} F^{{\rm c} Mv}
 -\frac{1}{r^2f(r)} \left( \nabla_M F^{{\rm a} Mr} + \epsilon^{\rm abc} C_M^{\rm b} F^{{\rm c} Mr} \right)=0. \label{EOM_dynamical_EF}
\end{align}
The constraint component of bulk EOMs is
\begin{align}
\nabla_M F^{{\rm a} Mr} + \epsilon^{\rm abc} C_M^{\rm b} F^{{\rm c} Mr}=0. \label{EOM_constraint_EF}
\end{align}
As emphasized below \eqref{Seff_formula}, we will solve the bulk dynamics in a partially on-shell approach: the profile of bulk gauge potential $C_\mu^{\rm a}$ will be fixed by solving dynamical components of bulk EOMs, say \eqref{EOM_dynamical_EF}. The constraint component of bulk EOMs, say \eqref{EOM_constraint_EF}, will be left aside.

Finally, the AdS boundary conditions are
\begin{align}
C_\mu(r=\infty_s,x^\mu) = \mathcal U(\varphi_s) (A_{s\mu} + i \partial_\mu ) \mathcal U^{\dagger}(\varphi_s) \equiv B_{s\mu}(x^\alpha), \quad \mathcal U(\varphi_s) = e^{i \varphi_s^{\rm a}(x) t^{\rm a}}, \quad s=1,2, \label{AdS_boundary_SU2}
\end{align}
where $\varphi_s^{\rm a}$ is the hydrodynamic field associated with SU(2) isospin charge, and $A_{s\mu}$ is the external (background) SU(2) gauge field in the boundary theory. For brevity, in \eqref{AdS_boundary_SU2} we dropped the prime in $A_{s\mu}^\prime$.

However, as explained in \cite{Glorioso:2018mmw} (see also \cite{Bu:2020jfo}), the AdS boundary conditions \eqref{AdS_boundary_SU2} are insufficient to fully determine $C_\mu$. Resultantly, we have freedom to impose
\begin{align}
C_v^{\rm a} (r=r_h-\epsilon, x^\mu) =0. \label{horizon_condition}
\end{align}
For a bulk U(1) field, the condition \eqref{horizon_condition} has been shown to be equivalent to discontinuity of $\partial_r C_v$ at the horizon \cite{Bu:2020jfo}. We will not elaborate on this point since linearized solution for $C_M^{\rm a}$ (which is identical to U(1) case) is sufficient for our purpose.

\section{From bulk dynamics to boundary effective action} \label{bulk_solution}

In this section we solve classical dynamics of the bulk theory, and then compute the partially on-shell bulk action $S_0[C_\mu[\varphi],\varphi]|_{\rm p.o.s}$ of \eqref{Seff_formula}.

\subsection{General consideration: linearize Yang-Mills system} \label{YM_linearize}

The dynamical EOMs \eqref{EOM_dynamical_EF} are nonlinear PDEs and are thus hard to solve analytically. Instead of solving \eqref{EOM_dynamical_EF} order-by-order in the boundary derivative expansion, we linearize the bulk Yang-Mills theory:
\begin{align}
& C_M^{\rm a} = \xi^1 C_M^{{\rm a}(1)} + \xi^2 C_M^{{\rm a}(2)} + \xi^3 C_M^{{\rm a}(3)} + \cdots, \nonumber
& F_{MN}^{\rm a} = \xi^1 F_{MN}^{{\rm a}(1)} + \xi^2 F_{MN}^{{\rm a}(2)} + \xi^3 F_{MN}^{{\rm a} (3)} + \cdots,
\end{align}
where
\begin{align}
& F_{MN}^{{\rm a}(1)} = \nabla_M C_N^{{\rm a}(1)} - \nabla_N C_M^{{\rm a}(1)}, \nonumber \\
& F_{MN}^{{\rm a}(2)} = \nabla_M C_N^{{\rm a}(2)} - \nabla_N C_M^{{\rm a}(2)} + \epsilon^{\rm abc} C_M^{{\rm b}(1)} C_N^{{\rm c}(1)} , \nonumber \\
& F_{MN}^{{\rm a}(3)} = \nabla_M C_N^{{\rm a}(3)} - \nabla_N C_M^{{\rm a}(3)} + \epsilon^{\rm abc} \left( C_M^{{\rm b}(2)} C_N^{{\rm c}(1)} + C_M^{{\rm b}(1)} C_N^{{\rm c}(2)} \right).
\end{align}
At each order in $\xi$-expansion, the dynamical EOMs \eqref{EOM_dynamical_EF} become a system of {\it linear} PDEs. For instance, at $\mathcal{O}(\xi^1)$ and $\mathcal{O}(\xi^2)$, we have
\begin{align}
\nabla_M F^{{\rm a} Mi (1)} =0, \qquad \qquad  \nabla_M F^{{\rm a} Mv (1)}
 -\frac{1}{r^2f(r)} \nabla_M F^{{\rm a} Mr (1)} = 0, \label{dynamical_EOM_1st}
\end{align}
\begin{align}
&\nabla_M F^{{\rm a} Mi (2)} + \epsilon^{\rm abc} C_M^{{\rm b}(1)} F^{{\rm c}Mi(1)}=0,  \nonumber \\
& \nabla_M F^{{\rm a} Mv (2)} + \epsilon^{\rm abc} C_M^{\rm b (1)} F^{{\rm c} Mv (1)}
 -\frac{1}{r^2f(r)} \left( \nabla_M F^{{\rm a} Mr (2)} + \epsilon^{\rm abc} C_M^{\rm b (1)} F^{{\rm c} Mr (1)} \right)=0. \label{dynamical_EOM_2nd}
\end{align}
The detailed forms of \eqref{dynamical_EOM_1st} and \eqref{dynamical_EOM_2nd} can be found in appendix \ref{YM_eom_explicit}.
The AdS boundary conditions \eqref{AdS_boundary_SU2} will be imposed as follows:
\begin{align}
C_{s\mu}^{{\rm a}(1)}(r= \infty_s, x^\alpha) = B_{s\mu}^{\rm a}(x^\alpha), \qquad   C_\mu^{{\rm a}(n)} (r= \infty_s, x^\alpha) = 0, \quad  n=2,3,\cdots, \quad s=1,2. \label{AdS_boundary_SU2_perturb}
\end{align}
Meanwhile, the vanishing horizon condition \eqref{horizon_condition} is also imposed perturbatively
\begin{align}
C_v^{{\rm a}(n)}(r_h - \epsilon, x^\mu) =0, \qquad n=1,2,3,\cdots. \label{horizon_condition_perturb}
\end{align}

The leading order EOMs \eqref{dynamical_EOM_1st} are homogeneous PDEs, which are identical to those of a free Maxwell field in Schwarzschild-AdS$_5$ geometry. Thus, the leading order solution $C_M^{{\rm a}(1)}$ is essentially the same as that constructed in \cite{Bu:2020jfo}, which will be reviewed in subsection \ref{review_LO_solution}.

The next-to-leading order EOMs \eqref{dynamical_EOM_2nd} form a system of linear inhomogeneous PDEs, which differ from leading order ones \eqref{dynamical_EOM_1st} by source terms, see appendix \ref{YM_eom_explicit} for more details. Moreover, the source terms are constructed from leading order solution $C_M^{{\rm a}(1)}$. Given that linearly independent solutions for homogeneous parts of \eqref{dynamical_EOM_2nd} have been worked out in \cite{Bu:2020jfo}, the task of solving \eqref{dynamical_EOM_2nd} boils down to looking for a particular solution associated with its source term. This can be implemented via Green's function approach \cite{Son:2009vu,Bu:2021clf}, which will be discussed in detail in appendix \ref{generic_structure_solution}.

We turn to the Yang-Mills action \eqref{S0}, which is expanded as:
\begin{align}
S_0 = \xi^2 S_0^{(2)} + \xi^3 S_0^{(3)} + \xi^4 S_0^{(4)}+ \cdots, \label{S0_expansion}
\end{align}
where
\begin{align}
& S_0^{(2)} = - \frac{1}{4} \int d^5x \sqrt{-g} F_{MN}^{{\rm a}(1)} F^{{\rm a} MN(1)}, \nonumber \\
& S_0^{(3)} = - \frac{1}{4} \int d^5x \sqrt{-g}\, 2 F_{MN}^{{\rm a}(1)} F^{{\rm a}MN(2)}, \nonumber \\
& S_0^{(4)} = - \frac{1}{4} \int d^5x \sqrt{-g} \left[ 2F_{MN}^{{\rm a}(1)} F^{{\rm a} MN(3)} + F_{MN}^{{\rm a}(2)} F^{{\rm a} MN(2)} \right]. \label{S0_expansion1}
\end{align}
The quadratic action $S_0^{(2)}$ will be similar to that of \cite{Bu:2020jfo}. The cubic action $S_0^{(3)}$ could be further simplified via integration by part:
\begin{align}
S_0^{(3)} & = - \frac{1}{2} \int d^5x \sqrt{-g} \left[ (\nabla_M C_N^{{\rm a}(2)} - \nabla_N C_M^{{\rm a} (2)}) F^{{\rm a} MN(1)} + \epsilon^{\rm abc} F_{MN}^{{\rm a}(1)} C^{{\rm b} M(1)} C^{{\rm c} N(1)} \right] \nonumber \\
& = - \frac{1}{2} \int d^5x \sqrt{-g} \left[ 2 \nabla_M C_N^{{\rm a}(2)} F^{{\rm a} MN(1)} + \epsilon^{\rm abc} F_{MN}^{{\rm a}(1)} C^{{\rm b} M(1)} C^{{\rm c} N(1)} \right] \nonumber \\
& = - \frac{1}{2} \int d^5x \sqrt{-g} \left[ 2 \nabla_M (C_N^{{\rm a}(2)} F^{{\rm a} MN(1)}) - C_N^{{\rm a}(2)} \nabla_M F^{{\rm a} MN(1)} + \epsilon^{\rm abc} F_{MN}^{{\rm a}(1)} C^{{\rm b} M(1)} C^{{\rm c} N(1)} \right] \nonumber \\
& = - \frac{1}{2} \int d^5x \sqrt{-g} \epsilon^{\rm abc} F_{MN}^{{\rm a}(1)} C^{{\rm b} M(1)} C^{{\rm c} N(1)}. \label{S0_cubic}
\end{align}
Explicitly, using the gauge-fixing \eqref{radial_gauge_Schw}, the cubic order action becomes
\begin{align}
S_0^{(3)} = - \int d^5x \sqrt{-g} \epsilon^{\rm abc} \left\{ (\partial_i C_v^{{\rm a}(1)} - \partial_v C_i^{{\rm a}(1)}) \frac{C_v^{{\rm b}(1)}}{r^4f(r)} C_i^{{\rm c}(1)} + \frac{1}{r^4} \partial_i C_j^{{\rm a}(1)} C_i^{{\rm b}(1)} C_j^{{\rm c} (1)} \right\}. \label{S0_cubic1}
\end{align}
The quartic order action $S_0^{(4)}$ of \eqref{S0_expansion1} could be simplified in the same fashion \eqref{S0_cubic}.
Eventually, the quartic order action is cast into
\begin{align}
S_0^{(4)} & = - \frac{1}{4} \int d^5x \sqrt{-g} \left[2 \epsilon^{\rm abc} F_{MN}^{{\rm a}(1)} C^{{\rm b} M(2)} C^{{\rm c} N(1)} + \epsilon^{\rm abc} (\nabla_M C_N^{{\rm a}(2)} - \nabla_N C_M^{{\rm a}(2)}) C^{{\rm b} M(1)} C^{{\rm c} N(1)}   \right. \nonumber \\
& \qquad \qquad \qquad \qquad \left. + C_M^{{\rm b}(1)} C^{{\rm b}M (1)} C_N^{{\rm c}(1)} C^{{\rm c}N (1)} - C_M^{{\rm b} (1)} C^{{\rm b} N(1)} C_N^{{\rm c}(1)} C^{{\rm c} M(1)} \right],
\end{align}
where the first two terms involve the next-to-leading order solution $C_M^{{\rm a}(2)}$, and the last two terms do not. Explicitly, the quartic order action is
\begin{align}
S_0^{(4)} = & - \frac{1}{4} \int d^5x \sqrt{-g} \left\{ - \frac{2}{r^4f(r)} C_v^{{\rm a}(1)} C_k^{{\rm b}(1)} \left( C_v^{{\rm a}(1)} C_k^{{\rm b}(1)} - C_k^{{\rm a}(1)} C_v^{{\rm b}(1)} \right) \right. \nonumber \\
& \qquad \qquad \qquad \qquad \left. + \frac{1}{r^4} C_k^{{\rm a}(1)} C_l^{{\rm b}(1)} \left( C_k^{{\rm a}(1)} C_l^{{\rm b}(1)} - C_l^{{\rm a}(1)} C_k^{{\rm b}(1)} \right)\right\} \nonumber \\
& - \int d^5x \sqrt{-g} \epsilon^{\rm abc} \left\{(\partial_i C_v^{{\rm a}(1)} - \partial_v C_i^{{\rm a}(1)}) \frac{C_v^{{\rm b}(2)}}{r^4f(r)} C_i^{{\rm c}(1)} + \frac{1}{r^4} \partial_i C_j^{{\rm a}(1)} C_i^{{\rm b}(2)} C_j^{{\rm c}(1)} \right\} \nonumber \\
& - \frac{1}{2}\int d^5x \sqrt{-g} \epsilon^{\rm abc} \left\{(\partial_i C_v^{{\rm a}(2)} - \partial_v C_i^{{\rm a}(2)}) \frac{C_v^{{\rm b}(1)}}{r^4f(r)} C_i^{{\rm c}(1)} + \frac{1}{r^4} \partial_i C_j^{{\rm a}(2)} C_i^{{\rm b}(2)} C_j^{{\rm c}(1)} \right\}. \label{S0_quartic}
\end{align}

In obtaining \eqref{S0_cubic1} and \eqref{S0_quartic}, we have imposed the leading-order dynamical EOMs \eqref{dynamical_EOM_1st}, but we did not impose the constraint equation \eqref{EOM_constraint_EF}. Moreover, we have utilized the following two facts: $C_\mu^{{\rm a}(2)}$ vanishes at the AdS boundaries, see \eqref{AdS_boundary_SU2_perturb}; leading order counterpart of the radial gauge choice \eqref{radial_gauge_Schw}.

\begin{figure}[htbp]
\centering
\includegraphics[width=0.9\textwidth]{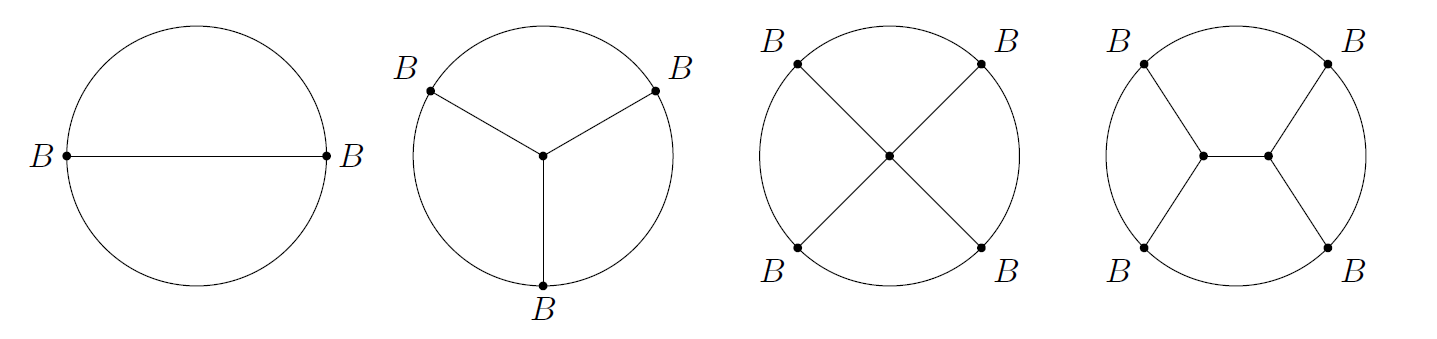}
\caption{Witten diagrams for $S_0^{(2)}$ (first), $S_0^{(3)}$ (second) and $S_0^{(4)}$ (last two). Here, the circle represents the AdS boundaries, and the region inside the circle is for AdS interior. The straight lines stand for bulk-to-bulk propagator (connecting two points inside the circle) and bulk-to-boundary propagator (connecting an interior point and a point on the circle) for bulk SU(2) gauge field propagating in doubled Schwarzschild-AdS.}
\label{witten_diagram}
\end{figure}

While the quadratic action $S_0^{(2)}$ and cubic action $S_0^{(3)}$ involve only the leading-order solution $C_\mu^{{\rm a}(1)}$, the computation of $S_0^{(4)}$ generally requires the next-to-leading order solution $C_\mu^{{\rm a}(2)}$. In subsection \ref{pos_solution}, we will see that the leading-order solution $C_\mu^{(1)}$ is essentially the bulk-to-boundary propagator. In appendix \ref{generic_structure_solution}, $C_\mu^{{\rm a}(2)}$ is constructed via Green's function method and would be schematically written as
\begin{align}
C_\mu^{{\rm a}(2)} \simeq \int_{\infty_2}^{\infty_1} dr^\prime G_{\perp, \parallel}(r,r^\prime, k^\alpha) \mathbb S_\mu^{{\rm a}(2)}(r^\prime,k^\alpha),
\end{align}
where $G_{\perp, \parallel}(r,r^\prime, k^\alpha)$ represents the bulk-to-bulk propagator, and the source term $\mathbb S_\mu^{{\rm a}(2)}(r^\prime,k^\alpha)$ are quadratic in $C_\mu^{{\rm a}(1)}$, see appendix \ref{generic_structure_solution}. Diagrammatically, the perturbative expansion of bulk action can be drawn as tree-level Witten diagrams, see Figure \ref{witten_diagram}. More precisely, the quadratic action $S_0^{(2)}$ and cubic action $S_0^{(3)}$ correspond to contact Witten diagrams, while the quartic action $S_0^{(4)}$ contains both contact Witten diagram (the first two lines of \eqref{S0_quartic}) and exchange Witten diagram (the last two lines of \eqref{S0_quartic}).

From subsection \ref{pos_solution}, it will be clear that $C_M^{{\rm a}(2)}$ starts from first order in boundary derivative expansion. This implies, through \eqref{S0_quartic}, that in practice we do not need to solve for $C_M^{{\rm a}(2)}$, since in present work we will truncate quartic order action $S_0^{(4)}$ at first order in the hydrodynamic derivative expansion. Equivalently, we will only compute contact-type Witten diagrams (the first three of Figure \ref{witten_diagram}).

\subsection{Partially on-shell solution: generic structure} \label{pos_solution}

Turning to Fourier space:
\begin{align}
C_M^{{\rm a}(n)}(r,x^\mu) = \int \frac{d\omega dq}{(2\pi)^2} e^{i k^\alpha x_\alpha} C_M^{{\rm a}(n)}(r,k^\mu), \quad n=1,2,\cdots, \quad k^\mu = (\omega,q,0,0),
\end{align}
we further reduce the system of linear PDEs, say \eqref{dynamical_EOM_1st} or \eqref{dynamical_EOM_2nd}, into a system of linear ordinary differential equations (ODEs). Here, by spatially rotational invariance, we have chosen the spatial momentum for Fourier modes to be along $x$-direction, without losing generality. Then, the bulk gauge field $C_\mu^{{\rm a}(n)}$ can be classified into two decoupled sub-sectors: the transverse sector $C_\perp^{{\rm a} (n)} $ with $\perp = y,z$ versus the longitudinal sector $C_\parallel^{{\rm a}(n)} = \left\{ C_v^{{\rm a} (n)}, ~ C_x^{{\rm a} (n)} \right\}$.

The transverse mode obeys a closed second order ODE ($n=1,2,\cdots$):
\begin{align}
\partial_r \left[ r^3f(r)\partial_r C_\perp^{{\rm a}(n)} \right] + \Box_\perp \left(\partial_r; \omega, q^2 \right) C_\perp^{{\rm a}(n)} = \mathbb S_\perp^{{\rm a}(n)}, \label{trans_eom}
\end{align}
where the source $\mathbb S_\perp^{{\rm a}(n)}$ is built from lower order solutions, with $\mathbb S_\perp^{{\rm a}(1)} = 0$, see \eqref{Source_perturb_EF}. From \eqref{Cmu_perturb_eom_EF}, the symbol $\Box_\perp \left(\partial_r; \omega, q^2 \right) C_\perp^{{\rm a}(n)}$ stands for terms containing boundary derivatives of $C_\perp^{{\rm a}(n)}$, and contains at most one first order radial derivative of $C_\perp^{{\rm a}(n)}$.

The dynamics for the longitudinal sector is more involved, see \eqref{Cmu_perturb_eom_EF} and \eqref{Source_perturb_EF}. Schematically, the dynamical EOMs are ($n=1,2,\cdots$)
\begin{align}
&\partial_r \left[ r^3 \partial_r C_v^{{\rm a}(n)} \right] + \Box_v\left( \partial_r; \omega, q\right)C_\parallel^{{\rm a} (n)} = \mathbb S_v^{{\rm a} (n)}, \nonumber \\
&\partial_r \left[ r^3 f(r)\partial_r C_x^{{\rm a}(n)} \right] + \Box_x\left( \partial_r; \omega, q\right)C_\parallel^{{\rm a} (n)} = \mathbb S_x^{{\rm a} (n)}, \label{longit_eom}
\end{align}
where, similar to transverse sector, the symbols $\Box_v\left( \partial_r; \omega, q\right)C_\parallel^{{\rm a} (n)}$ and $\Box_x\left( \partial_r; \omega, q\right)C_\parallel^{{\rm a} (n)}$ denote terms having boundary derivatives of $C_{v,x}^{{\rm a}(n)}$, and contain at most one first order radial derivative of $C_{v,x}^{{\rm a}(n)}$. Finally, $\mathbb S_{v,x}^{{\rm a} (n)}$ are the source terms to be constructed from lower order solutions, with the leading order $\mathbb S_{v,x}^{{\rm a} (1)} =0$.

In general, solution for \eqref{trans_eom} or \eqref{longit_eom} consists of two parts: generic solution for homogeneous parts of \eqref{trans_eom} and \eqref{longit_eom}, which will be referred to as homogeneous generic solution (HGS), and particular solution for the whole inhomogeneous systems \eqref{trans_eom} and \eqref{longit_eom}. First, we consider HGS by ignoring source terms, which has been worked out in Section 4 of \cite{Bu:2020jfo}. We cut the radial contour of Figure \ref{holographic_SK_contour} at the leftmost point $r=r_h-\epsilon$ so that the contour is split into upper branch and lower branch. Then, we search for linearly independent solutions on single copy AdS, i.e., either on the upper branch or on the lower branch of the contour. For generic values of $\omega$ and $q$, these linearly independent solutions are not known analytically, and may be classified according to their near-horizon behavior. Resultantly, on upper (lower) branch of the radial contour, the HGS is simply linear superposition of those linearly independent solutions. With HGS at hand, it is straightforward to obtain particular solution for the whole inhomogeneous system, for instance, by Green's function approach. On upper (lower) branch, we sum the HGS and particular solution, forming inhomogeneous generic solution (IHGS). At the cutting slice $r=r_h - \epsilon$, the IHGS on the upper branch will be properly glued to the IHGS on the lower branch. Eventually, we impose AdS boundary conditions. This completes the solving of \eqref{trans_eom} and \eqref{longit_eom}.

Finally, we briefly comment on the next-to-leading order solution $C_\mu^{{\rm a}(2)}$. From \eqref{dynamical_EOM_2nd}, it is direct to check that the source terms $\mathbb S_{\perp,v,x}^{{\rm a}(2)}$ appearing in \eqref{trans_eom} and \eqref{longit_eom} are at least of first order in boundary derivative expansion. Given homogenous AdS boundary conditions for $C_\mu^{{\rm a}(2)}$, see \eqref{AdS_boundary_SU2_perturb}, one can immediately conclude that $C_\mu^{{\rm a}(2)}$ will be at least of first order in boundary derivative expansion, just as their source terms. Consequently, combined with \eqref{S0_quartic}, this observation implies that the contribution from $C_\mu^{{\rm a}(2)}$ will start from second order in boundary derivative expansion, which will be beyond focus of present work. Since the next-to-leading solution is irrelevant in subsequent calculations, we postpone the detailed construction of it in appendix \ref{generic_structure_solution}.

\subsection{Review of leading-order solution of \cite{Bu:2020jfo}} \label{review_LO_solution}

In this subsection, we review the leading-order solution comprehensively constructed in \cite{Bu:2020jfo}. The strategy of solving leading-order counterparts (i.e., with $n=1$) of \eqref{trans_eom} and \eqref{longit_eom} will go under three steps.

First, with the radial contour of Figure \ref{holographic_SK_contour} cut at $r=_h - \epsilon$, we search for linearly independent solutions when $r$ varies on the upper branch or lower branch. While this topic has been widely explored in the literature, a major difference arises due to partially on-shell approach adopted in this work. Particularly, for the longitudinal modes, one has to include a new linearly independent solution, referred to as ``polynomial solution'' in \cite{Bu:2020jfo}, which does not obey the constraint equation and is thus off-shell. Then, the generic solution on the upper (lower) branch is simply superposition of the linearly independent solutions.

Second, we will perform a gluing procedure so that the generic solution on the upper branch will be properly glued with that on the lower branch. Here, the gluing condition is derived from the requirement that variational problem for $S_0^{(2)}$ of \eqref{S0_expansion1} is well-defined at the cutting slice $r=r_h - \epsilon$, say $\delta S_0^{(2)}/\delta C_\mu^{{\rm a} (1)} \big|_{r=r_h -\epsilon} =0$. Essentially, the gluing condition tells how the radial derivatives of bulk fields will jump across the cutting slice $r= r_h - \epsilon$. Here, it is natural to assume that bulk fields $C_\mu^{{\rm a}(1)}$ are continuous across $r=r_h - \epsilon$. We refer the reader into \cite{Bu:2020jfo} for more details.

Finally, superposition constants in the generic solutions will be completely fixed by the doubled AdS boundary conditions.

Without repeating the details, here we simply write down final results for leading order solution $C_\mu^{{\rm a}(1)}$ \cite{Bu:2020jfo}. For the transverse mode $C_\perp^{{\rm a}(1)}$,
\begin{align}
& C_\perp^{{\rm a}(1)~{\rm up}}(r,k^\mu) = c_\perp(k^\mu) C_\perp^{\rm ig} (r,k^\mu) - h_\perp(k^\mu) C_\perp^{\rm ig}(r,\bar k^\mu) e^{2i\omega \zeta_2(r)}, ~\qquad \quad r\in [r_h-\epsilon, \infty_2), \nonumber \\
& C_\perp^{{\rm a}(1)~{\rm dw}}(r,k^\mu) = c_\perp(k^\mu) C_\perp^{\rm ig} (r,k^\mu) - h_\perp(k^\mu) e^{-\beta\omega} C_\perp^{\rm ig}(r,\bar k^\mu) e^{2i\omega \zeta_1(r)}, \quad r\in [r_h-\epsilon, \infty_1), \label{Cperp_solution_LO}
\end{align}
where $\bar k^\mu= (-\omega,q,0,0)$. $C_\perp^{\rm ig} (r,k^\mu)$ and $C_\perp^{\rm ig}(r,\bar k^\mu) e^{2i\omega \zeta_s(r)}$ represent the ingoing solution and outgoing solution\footnote{Here, as in \cite{Chakrabarty:2019aeu,Jana:2020vyx,Ghosh:2020lel,Bu:2020jfo}, the ingoing and outgoing modes are related via time-reversal symmetry.} for homogeneous part of \eqref{trans_eom}. The functions $\zeta_{1,2}(r)$ are
\begin{align}
\zeta_s(r) = \int_{\infty_s}^r \frac{dy}{y^2f(y)}, \qquad r \in[r_h-\epsilon, \infty_s), \quad s=1 ~{\rm or}~2,
\end{align}
which has an explicit form:
\begin{align}
\zeta_s(r) = - \frac{1}{4r_h} \left[ \pi - 2\arctan\left( \frac{r}{r_h} \right) + \log\left(1+\frac{r_h}{r}\right) - \log\left(1-\frac{r_h}{r}\right) \right].
\end{align}
The superposition coefficients $c_\perp, h_\perp$ are determined by the AdS boundary conditions (see \eqref{AdS_boundary_SU2_perturb}):
\begin{align}
c_\perp = \frac{1}{2} \coth\frac{\beta \omega}{2} \frac{B_{a\perp}^{\rm a}(k^\mu)} {C_\perp^{\rm ig(0)}(k^\mu)} + \frac{B_{r\perp}^{\rm a}(k^\mu)} {C_\perp^{\rm ig(0)}(k^\mu)},
 \qquad h_\perp = \frac{B_{a\perp}(k^\mu)}{(1-e^{-\beta \omega})C_\perp^{\rm ig(0)}(\bar k^\mu)}, \label{chperp}
\end{align}
where $C_\perp^{\rm ig(0)}(k^\mu)$ is the boundary value of ingoing mode $C_\perp^{\rm ig} (r,k^\mu)$. In ingoing EF coordinate system, the ingoing solution $C_\perp^{\rm ig}(r,k^\mu)$ is regular over the entire contour, particularly near the horizon.
Here, we introduced the $(r,a)$-basis:
\begin{align}
B_{r\mu}^{\rm a} = \frac{1}{2}\left( B_{1\mu}^{\rm a} + B_{2\mu}^{\rm a} \right), \qquad \qquad B_{a\mu}^{\rm a} = B_{1\mu}^{\rm a} - B_{2\mu}^{\rm a}, \qquad \mu = v, x, \perp.
\end{align}

For the longitudinal modes, we have
\begin{align}
C_v^{{\rm a}(1)~{\rm up}}(r,k^\mu) = & c_\parallel(k^\mu) C_v^{\rm ig}(r,k^\mu) + h_\parallel(k^\mu) C_v^{\rm ig}(r,\bar k^\mu) e^{2i\omega\zeta_2(r)} + p_\parallel^{\rm up} C_v^{\rm pg}(r,k^\mu) \nonumber \\
& + n_\parallel^{\rm up}(k^\mu) C_v^{\rm pn}(r,k^\mu), \nonumber \\
C_x^{{\rm a}(1)~{\rm up}}(r,k^\mu) = & c_\parallel(k^\mu) C_x^{\rm ig}(r,k^\mu) - h_\parallel(k^\mu) C_x^{\rm ig}(r,\bar k^\mu) e^{2i\omega \zeta_2(r)} + p_\parallel^{\rm up} C_v^{\rm pg}(r,k^\mu)  \nonumber \\
& + n_\parallel^{\rm up}(k^\mu) C_x^{\rm pn}(r,k^\mu), \nonumber \\
C_v^{{\rm a}(1)~{\rm dw}}(r,k^\mu) = &c_\parallel(k^\mu) C_v^{\rm ig}(r,k^\mu) + h_\parallel(k^\mu) e^{-\beta \omega} C_v^{\rm ig}(r,\bar k^\mu) e^{2i\omega\zeta_1(r)} + p_\parallel^{\rm dw} C_x^{\rm pg}(r,k^\mu) \nonumber \\
& + n_\parallel^{\rm up}(k^\mu) C_v^{\rm pn}(r,k^\mu), \nonumber \\
C_x^{{\rm a}(1)~{\rm dw}}(r,k^\mu) = & c_\parallel(k^\mu) C_x^{\rm ig}(r,k^\mu) - h_\parallel(k^\mu) e^{-\beta \omega} C_x^{\rm ig}(r,\bar k^\mu) e^{2i\omega\zeta_1(r)}+ p_\parallel^{\rm dw} C_x^{\rm pg}(r,k^\mu) \nonumber \\
& + n_\parallel^{\rm dw}(k^\mu) C_x^{\rm pn}(r,k^\mu), \label{Cparallel_solution_LO}
\end{align}
where, as for the transverse mode, $C_{v,x}^{\rm ig}$ and $C_{v,x}^{\rm ig}e^{2i\omega\zeta_s(r)}$ are the ingoing and outgoing solutions, respectively. $C_{v,x}^{\rm pg}$ is the pure gauge solutions, which is obtained from a zero solution by a gauge transformation preserving the gauge condition \eqref{radial_gauge_Schw}. Lastly, $C_{v,x}^{\rm pn}$ represents the off-shell solution which is crucial in exhausting the full basic solutions. Among the four linearly independent solutions, only $C_v^{\rm pg}$ does not vanish at the horizon. Consequently, the condition \eqref{horizon_condition_perturb} requires to set
\begin{align}
p_\parallel^{\rm up} = p_\parallel^{\rm dw}=0.
\end{align}
The rest superposition coefficients $c_\parallel, h_\parallel, n_\parallel^{\rm up}, n_\parallel^{\rm dw}$ are determined by the AdS boundary conditions \eqref{AdS_boundary_SU2_perturb}:
\begin{align}
& c_\parallel= \frac{1}{2}G_1^{-1} \left\{ 2B_{rx}^{\rm a}(k^\mu) C_v^{\rm pn(0)}(k^\mu) - 2 B_{rv}^{\rm a}(k^\mu) C_x^{\rm pn(0)}(k^\mu) \right. \nonumber\\
& \qquad \qquad \left.+ \coth\frac{\beta \omega}{2} \left[B_{ax}^{\rm a}(\omega,q) C_v^{\rm pn(0)}(k^\mu) - B_{av}^{\rm a}(k^\mu) C_x^{\rm pn(0)} (k^\mu) \right] \right\}, \label{c_sol}\\
& h_\parallel = (1-e^{-\beta \omega})^{-1} G_2^{-1} \left[ B_{ax}^{\rm a}(k^\mu) C_v^{\rm pn(0)} (k^\mu) - B_{av}^{\rm a}(k^\mu) C_x^{\rm pn(0)}(k^\mu) \right], \label{h_sol}\\
& n_\parallel^{\rm dw}- n_\parallel^{\rm up}= G_2^{-1} \left[B_{ax}^{\rm a}(k^\mu) C_v^{\rm ig(0)}(\bar k^\mu) + B_{av}^{\rm a}(k^\mu) C_x^{\rm ig(0)} (\bar k^\mu) \right], \label{n-_sol}\\
& \frac{1}{2}(n_\parallel^{\rm dw} + n_\parallel^{\rm up}) = - G_1^{-1} \left[ B_{rx}^{\rm a}(k^\mu) C_v^{\rm ig(0)}(k^\mu) - B_{rv}^{\rm a}(k^\mu) C_x^{\rm ig(0)} (k^\mu) \right] - \frac{1}{2}\coth\frac{\beta \omega}{2}\nonumber \\
& \qquad \qquad \qquad \quad \times   G_1^{-1} G_2^{-1} G_3 \left[  B_{ax}^{\rm a}(k^\mu) C_v^{\rm pn(0)}(k^\mu) - B_{av}^{\rm a}(k^\mu) C_x^{\rm pn(0)} (k^\mu) \right], \label{n+_sol}
\end{align}
where
\begin{align}
& G_1=C_v^{\rm pn(0)}(k^\mu) C_x^{\rm ig(0)}(k^\mu)- C_v^{\rm ig(0)}(k^\mu) C_x^{\rm pn(0)}(k^\mu), \nonumber \\
& G_2=C_v^{\rm pn(0)}(k^\mu) C_x^{\rm ig(0)}(\bar k^\mu) + C_v^{\rm ig(0)}(\bar k^\mu) C_x^{\rm pn(0)}(k^\mu), \nonumber \\
& G_3= C_v^{\rm ig(0)} (\bar k^\mu) C_x^{\rm ig(0)}(k^\mu) + C_v^{\rm ig(0)}(k^\mu) C_x^{\rm ig(0)}(\bar k^\mu).
\end{align}
Here, we use the superscript $^{(0)}$ to denote boundary values of various linearly independent solutions. It is not difficult to verify that $G_1 \neq 0$.

Practically, these linearly independent solutions were computed \cite{Bu:2020jfo} in Schwarzschild coordinate system, which can be converted to EF coordinate system through:
\begin{align}
& C_\perp^{\rm ig}(r,k^\mu)= \tilde C_\perp^{\rm ig}(r,k^\mu) e^{i\omega \zeta_s(r)}, ~\qquad C_\parallel^{\rm ig}(r,k^\mu)= \tilde C_\parallel^{\rm ig}(r,k^\mu) e^{i\omega \zeta_s(r)}, \nonumber \\
& C_\parallel^{\rm pg}(r,k^\mu)= \tilde C_\parallel^{\rm pg}(r,k^\mu) e^{i\omega \zeta_s(r)}, \qquad C_\parallel^{\rm pn}(r,k^\mu)= \tilde C_\parallel^{\rm pn}(r,k^\mu) e^{i\omega \zeta_s(r)},
\end{align}
where $s=1$ when $r\in [r_h-\epsilon, \infty_1)$ and $s=2$ when $r\in [r_h - \epsilon, \infty_2)$. Here, the tilded functions denote linearly independent solutions in the Schwarzschild coordinate system.

For later calculations, we summarize the hydrodynamic expansion for all linearly independent solutions when $\omega, q \ll T$. It is convenient to introduce a new radial coordinate $u$:
\begin{align}
u=r_h^2/r^2 \Longrightarrow \tilde C_\mu(r,\omega,q) \to \tilde C_\mu(u,\omega,q). \label{u_r}
\end{align}
For the transverse mode, we just need the ingoing solution
\begin{align}
\tilde C_\perp^{\rm ig}(u,\omega,q)= (1-u^2)^{-i\tilde \omega/2} \left\{ 1+ i\tilde \omega \log(1+u) + \frac{1}{24} \pi^2(3\tilde \omega^2- 2\tilde q^2) - \frac{1}{4}\tilde\omega^2 \log^22 \right. \nonumber \\
 \left. + \frac{1}{2} \tilde \omega^2 \log(1-u) \log \frac{2}{1+u} - \frac{1}{4} \log(1+u) \left[2(\tilde\omega^2- \tilde q^2) \log u + \tilde \omega^2 \log(1+u) \right] \right. \nonumber \\
\left. +\frac{1}{2} (\tilde q^2- \tilde \omega^2) \left[{\rm Li}_2(1-u) + {\rm Li}_2(-u) \right] - \frac{1}{2} \tilde \omega^2 {\rm Li}_2\left(\frac{1+u}{2}\right) + \cdots \right\},  \label{Cperp_ingoing_hydro}
\end{align}
where $\rm Li_2$ is the Polylogarithm function, and the tilded frequency and momentum are dimensionless
\begin{align}
\tilde \omega \equiv \frac{\omega}{2r_h} = \frac{\omega}{2\pi T}, \qquad \tilde q \equiv \frac{q}{2r_h} = \frac{q}{2\pi T}.
\end{align}
For the longitudinal sector, the ingoing solution and polynomial solution are
\begin{align}
\tilde C_t^{\rm ig}(u, \omega,q)&= (1-u^2)^{1-i\tilde \omega/2} \left[\frac{i\tilde q } {1+u} +\frac{\tilde \omega \tilde q }{1-u^2} \left( \log\frac{2}{1+u} + u\log u \right)  +\cdots\right], \nonumber \\
\tilde C_x^{\rm ig}(u,\omega,q)&= (1-u^2)^{-i\tilde \omega/2} \left\{1 + i\tilde \omega \log\frac{1+u}{2} + \frac{\pi^2 \tilde \omega^2}{24} - \frac{1}{2} \tilde\omega^2 \log\frac{1-u}{2} \log \frac{1+u}{2} \right. \nonumber \\
& \left. - \frac{1}{4}\tilde\omega^2 \log^2\frac{1+u}{2} - \frac{1}{2}\tilde \omega^2 \log u \log(1+u) - \frac{1}{2} \tilde \omega^2 {\rm Li}_2(1-u) - \frac{1}{2} \tilde \omega^2 {\rm Li}_2(-u)\right. \nonumber \\
& \left.- \frac{1}{2} \tilde \omega^2 {\rm Li}_2\left( \frac{1+u}{2}\right) +\cdots \right\}. \label{Cparallel_ingoing_hydro}
\end{align}
\begin{align}
& \tilde C_t^{\rm pn}(u,\omega,q) = 2(1-u) + 2\tilde q^2 \left[u\log u +(1+u) \log \frac{2}{1+u} \right] + \cdots , \nonumber \\
& \tilde C_x^{\rm pn}(u,\omega,q) = -\tilde \omega \tilde q \left\{\frac{1}{4}(\pi^2 - 2\log^22) + \log u \log\frac{1+u}{1-u} + \frac{1}{2} \log(1+u) \log\frac{4}{1+u} \right. \nonumber \\
& \qquad \qquad \qquad \qquad \quad \left. - {\rm Li}_2\left(\frac{1-u}{2}\right) + {\rm Li}_2(-u) - {\rm Li}_2(u) \right\} + \cdots.  \label{Cparallel_polynomial_hydro}
\end{align}
Given that the two limits $\epsilon \to 0$ and $\tilde \omega \to 0$ do not commute \cite{Glorioso:2018mmw,Bu:2020jfo}:
\begin{align}
\lim_{\epsilon\to 0} \lim_{\tilde \omega \to 0} (1-u^2)^{-i\tilde \omega/2} \neq \lim_{\tilde \omega \to 0} \lim_{\epsilon\to 0} (1-u^2)^{-i\tilde \omega/2}, \qquad {\rm as}~~ u \to 1-\epsilon, \label{subtelty1}
\end{align}
in \eqref{Cperp_ingoing_hydro}, \eqref{Cparallel_ingoing_hydro} and \eqref{Cparallel_polynomial_hydro}, we have kept the overall oscillating factor like $(1-u^2)^{-i\tilde \omega/2}$ (if present) unexpanded in small $\tilde \omega$. Under this treatment, the leading order solutions \eqref{Cperp_solution_LO} and \eqref{Cparallel_solution_LO} would be schematically written as
\begin{align}
& C_\perp^{{\rm a}(1)} = \mathcal C_\perp(r,\omega,q) + [f(r)]^{i\tilde \omega} \mathcal H_\perp(r,\omega,q), \nonumber \\
& C_\parallel^{{\rm a}(1)} = \mathcal C_\parallel(r,\omega,q) + [f(r)]^{i\tilde \omega} \mathcal H_\parallel(r,\omega,q) + [f(r)]^{i\tilde \omega/2} \mathcal G_\parallel(r,\omega,q), \label{LO_soluton_schematic}
\end{align}
where $\mathcal C_{\perp,\parallel}$, $\mathcal H_{\perp, \parallel}$, and $\mathcal G_\parallel$ represent regular parts of the leading order solutions, which are valid to be expanded in terms of $\tilde \omega, \tilde q$ even near the horizon. Essentially, the results presented in \eqref{Cperp_ingoing_hydro}, \eqref{Cparallel_ingoing_hydro} and \eqref{Cparallel_polynomial_hydro} correspond to hydrodynamic expansion of $\mathcal C_{\perp,\parallel}$, $\mathcal H_{\perp, \parallel}$, and $\mathcal G_\parallel$.

\subsection{Boundary effective action: contour integrals} \label{boundary_action}

In this subsection, we compute the radial contour integrals in the bulk action \eqref{S0_expansion}, producing the boundary effective action \eqref{Seff_formula}, which would be expanded in amplitude of $B_{r\mu}$ and $B_{a\mu}$,
\begin{align}
S_{eff} = S_{eff}^{(2)} + S_{eff}^{(3)} + S_{eff}^{(4)} + \cdots
= \int d^4x \left[ \mathcal L_{eff}^{(2)} + \mathcal L_{eff}^{(3)} + \mathcal L_{eff}^{(4)} + \cdots \right]. \label{Seff_expansion}
\end{align}
At each order in amplitude expansion \eqref{Seff_expansion}, we will truncate the action at first order in hydrodynamic derivative expansion.

Near the two AdS boundaries $r=\infty_1$ and $r=\infty_2$, the contour integrals in $S_0^{(2)}$, $S_0^{(3)}$ and $S_0^{(4)}$ suffer from UV divergences, which are exactly cancelled by the counter-term action like \eqref{Sct}, added on each AdS boundary:
\begin{align}
S_{\rm ct} = \frac{1}{4} \log r \, \int d^4x \sqrt{-\gamma} F_{\mu\nu}^{\rm a} F^{{\rm a} \mu\nu}\bigg|_{r= \infty_1} - \frac{1}{4} \log r \, \int d^4x \sqrt{-\gamma} F_{\mu\nu}^{\rm a} F^{{\rm a} \mu\nu}\bigg|_{r= \infty_2}
\end{align}
where the extra minus sign for the second term is due to flipping of $r$-orientation.

~

$\bullet$ {\bf Quadratic Lagrangian $\mathcal L_{eff}^{(2)}$}

~

Through integration by part, the $S_0^{(2)}$ of \eqref{S0_expansion1} is reduced into a surface term \cite{Bu:2020jfo},
\begin{align}
S_0^{(2)} = - \frac{1}{2} \int d^4x \sqrt{-\gamma} n_M C_N^{\rm a} F^{{\rm a} MN} \bigg|_{r=\infty_2}^{r=\infty_1}.
\end{align}
which helps to avoid contour integrals. Up to first order in boundary derivative expansion, the quadratic action $S_{eff}^{(2)}$ is \cite{Bu:2020jfo,Glorioso:2018mmw,deBoer:2018qqm}
\begin{align}
\mathcal L_{eff}^{(2)} = & \frac{i}{2} B_{av}^{\rm a} w_1 B_{av}^{\rm a} + \frac{i}{2} B_{ak}^{\rm a} w_2 B_{ak}^{\rm a} + i B_{av}^{\rm a} w_4 \partial_k B_{ak}^{\rm a} + B_{av}^{\rm a} w_5 B_{rv}^{\rm a} \nonumber \\
&+ \partial_k B_{ak}^{\rm a} w_7 B_{rv}^{\rm a} + B_{ak}^{\rm a} w_8 \partial_v B_{rk}^{\rm a},  \label{Leff_quadratic}
\end{align}
where
\begin{align}
& w_1 = 0+ \mathcal{O}(\partial^2), ~ ~ ~\qquad w_2= \frac{2r_h^2}{\pi} + \mathcal{O}(\partial^2), \qquad  w_4 = 0+ \mathcal{O}(\partial^1), \nonumber \\
& w_5 = 2r_h^2 + \mathcal{O}(\partial^2), \qquad w_7 = 0 + \mathcal{O}(\partial^1), \qquad \quad w_8 = -r_h + \mathcal{O}(\partial^1). \label{w_expansion}
\end{align}

~

$\bullet$ {\bf Strategy of computing $\mathcal L_{eff}^{(3)}$ and $\mathcal L_{eff}^{(4)}$}

~

The computation of $S_0^{(3)}$ and $S_0^{(4)}$ inevitably involves contour integrals. Due to presence of branch cuts in the integrands, we advance by splitting the radial contour as
\begin{align}
\int_{\infty_2}^{\infty_1} dr = \int_{\infty_2}^{r_h+\epsilon} dr + \int_{\mathcal C} dr + \int_{r_h+\epsilon}^{\infty_1} dr, \label{contour_split}
\end{align}
where $\mathcal C$ denotes the infinitesimal circle. First, let us examine the near-horizon behavior for $S_0^{(3)}$ and $S_0^{(4)}$. Recall that the leading order solution $C_v^{{\rm a}(1)}$ vanishes at the horizon, and the regular part of the spatial component $C_i^{{\rm a}(1)}$ is finite near the horizon. Thus, it is obvious that the integrands in $S_0^{(3)}$ \eqref{S0_cubic1} and $S_0^{(4)}$ \eqref{S0_quartic} are finite when $r$ varies on the infinitesimal circle. Therefore, in both \eqref{S0_cubic1} and \eqref{S0_quartic} contributions from integrals along the infinitesimal circle will vanish as $\epsilon \to 0$ is taken in the end. Eventually, the contour integrals in \eqref{S0_cubic1} and \eqref{S0_quartic} reduce into real-variable integrals on the interval $[r_h +\epsilon, \infty)$, which we schematically write as,
\begin{align}
S_0^3 &= \int \frac{d^2k_1 d^2 k_2 d^2 k_3}{(2\pi)^6} \delta^{(2)} (k_1 + k_2 + k_3) \mathcal I(k_1,k_2,k_3) \nonumber \\
S_0^4 &= \int \frac{d^2k_1 d^2 k_2 d^2 k_3 d^2 k_4}{(2\pi)^8} \delta^{(2)} (k_1 + k_2 + k_3 + k_4) \mathcal J(k_1, k_2, k_3, k_4) \label{S0_cubic_quartic_real_integral}
\end{align}
with (for convenience we convert to $u$-variable by \eqref{u_r})
\begin{align}
\mathcal I &= \int_{r_h+ \epsilon}^\infty dr \sum_{m}(r-r_h)^{i \lambda_m} \mathcal I_m(r,k_1, k_2, k_3), \nonumber \\
\mathcal J &= \int_{r_h + \epsilon}^\infty dr \sum_m (r-r_h)^{i \delta_m} \mathcal J_m(r,k_1, k_2, k_3,k_4). \label{IJ_formula}
\end{align}
Here, $\mathcal I_m, \mathcal J_m$ are constructed from (essentially products of) regular parts of leading order solution, cf. \eqref{LO_soluton_schematic}. Importantly, $\mathcal I_m$ and $\mathcal J_m$ are regular functions, i.e., they does not contain singularity over the interval $[r_h, \infty)$. In \eqref{IJ_formula}, $\lambda_m$ is certain linear combination of $k_1^0, k_2^0, k_3^0$ while $\delta_m$ is certain linear combination of $k_1^0, k_2^0, k_3^0, k_4^0$, whose exact forms will be irrelevant in general analysis below.

In general, we have two different treatments in extracting hydrodynamic limits of \eqref{IJ_formula}. In {\bf Scheme I}, we will expand regular functions $\mathcal I_m, \mathcal J_m$ in terms of four-momentum, but will keep the oscillating factors $(r-r_h)^{i\lambda_m,\, i\delta_m}$ unexpanded; then, once the radial integrals in \eqref{IJ_formula} are done, we need another hydrodynamic expansion. In {\bf Scheme II}, we expand both oscillating factors $(1-u)^{i\lambda_m,\, i\delta_m}$ and regular parts $\mathcal I_m, \mathcal J_m$ in the hydrodynamic limit, and then perform the radial integrals \eqref{IJ_formula}. Generically, thanks to the subtlety \eqref{subtelty1}, the results obtained within these two schemes would not match. Nevertheless, we observe that up to first order in the boundary derivative expansion, these two schemes accidentally yield the same results. This observation relies on the fact that both $\mathcal I_m$ and $\mathcal J_M$ can be represented by Taylor series near the horizon
\begin{align}
\mathcal I_m = \sum_{n =0}^{\infty} \mathcal I_m^{(n)}(k_1,k_2,k_3) (r-r_h)^n, \qquad \mathcal J_m = \sum_{n =0}^{\infty} \mathcal J_m^{(n)}(k_1,k_2,k_3,k_4) (r-r_h)^n, \label{Taylor_expansion_IJ}
\end{align}
which is convergent over the whole interval $r \in [r_h,\infty]$. Then, the task of extracting hydrodynamic limits of $S_0^{(3)}$ and $S_0^{(4)}$ boils down to evaluating the following type integrals in the hydrodynamic limit
\begin{align}
\mathfrak{C}_n= \int dr (r-r_h)^{i\lambda_m} (r-r_h)^n
\end{align}
Direct calculation gives
\begin{align}
&\mathfrak{C}_n \big|_{\rm Scheme~I} = \sum_{l=0}^{\infty} \frac{(n+1)^{-l-1} (-i\lambda_m )^l \Gamma \left[ l+1,-(n+1) \log (r-r_h) \right]}{\Gamma (l+1)}, \nonumber \\
&\mathfrak{C}_n \big|_{\rm Scheme~II} = \frac{(r-r_h)^{i\lambda_m +n +1}}{i\lambda_m + n+1}
\end{align}
which can be shown to be equivalent in hydrodynamic limit. Here, $\Gamma [s]$ and $\Gamma[s,x]$ are the Euler gamma function and incomplete gamma function, respectively.

Therefore, in subsequent calculations, it is valid to simply expand $C_\mu^{{\rm a}(1)}$ in the hydrodynamic limit (including the oscillating factors).

With \eqref{Cperp_ingoing_hydro}, \eqref{Cparallel_ingoing_hydro}, and \eqref{Cparallel_polynomial_hydro}, we truncate the leading order solution \eqref{Cperp_solution_LO} and \eqref{Cparallel_solution_LO} to first order in $\tilde \omega$ and $\tilde q$:
\begin{align}
	&C_v^{{\rm a}(1)\, {\rm up}}(r,k^\mu)=\left[1+i \omega \zeta_2(r)\right] \left[2 B_{r v}^{\rm a} F_2(r) - B_{a v}^{\rm a} F_2(r)\right], \nonumber \\
	&C_v^{{\rm a}(1)\, {\rm dw}}(r,k^\mu)=\left[1+i \omega \zeta_1(r)\right]\left[ 2 B_{r v}^{\rm a} F_2(r) + B_{a v}^{\rm a} F_2(r)\right], \nonumber \\
	&C_i^{{\rm a}(1)\, {\rm up}}(r,k^\mu)=\left[1+i \omega \zeta_2(r)\right] \left[ B_{r i}^{\rm a} + B_{a i}^{\rm a}  \left(F_1(r)-\frac{1}{2}\right)\right] +\frac{1}{2} \pi \tilde\omega (2B_{r i}^{\rm a}+ B_{a i}^{\rm a}) F_1(r) , \nonumber \\
	& C_i^{{\rm a}(1)\, {\rm dw}}(r,k^\mu)=\left[1+i \omega \zeta_1(r)\right] \left[B_{r i}^{\rm a} + B_{a i}^{\rm a}  \left(F_1(r)+\frac{1}{2}\right)\right] +\frac{1}{2} \pi \tilde \omega (2B_{r i}^{\rm a}-B_{a i}^{\rm a}) F_1(r) , \label{LO_solution_expansion}
\end{align}
where $r\in [r_h+ \epsilon, \infty_2)$ for solutions on the upper branch and $r\in [r_h+\epsilon, \infty_1)$ for solutions on the lower branch. The functions $F_{1,2}(r)$ are
\begin{align}
F_1(r) = \frac{i}{2 \pi}\left[ \log \left(r^2+r_h^2\right)-\log \left(r^2-r_h^2\right) \right], \qquad \qquad F_2(r) = \frac{r^2-r_h^2}{2 r^2}.
\end{align}
From \eqref{S0_cubic1} and \eqref{S0_quartic}, the leading order solution $C_\mu^{{\rm a} (1)}$ contributes to $S_0^{(3)}$ and $S_0^{(4)}$ through itself or its boundary derivatives. Thus, to first order in boundary derivative, the contributions from $i\omega \zeta_s(r)$-terms in \eqref{LO_solution_expansion} to $S_0^{(3)}$ and $S_0^{(4)}$ will vanish, due to the delta-functions in \eqref{S0_cubic_quartic_real_integral}.

Plugging \eqref{LO_solution_expansion} into \eqref{S0_cubic1} and \eqref{S0_quartic}, we analytically implement the radial integrals. Below, we present the final results for $S_{eff}^{(3)}$ and $S_{eff}^{(4)}$.

~

$\bullet$ {\bf Leading order result of $\mathcal L_{eff}^{(3)}$}

~

The cubic effective Lagrangian $\mathcal L_{eff}^{(3)}$ starts from first order in boundary derivative expansion:
\begin{align}
\mathcal L_{eff}^{(3)} = & -4\lambda_1{\rm tr} \left[ B_{\{rv} B_{ri} \partial_i B_{a\}v}\right] -4\lambda_2 {\rm tr}\left[B_{ai} \partial_i B_{(av} B_{r)v}\right]
-4\lambda_3 {\rm tr} [B_{av} B_{ai} \partial_i B_{av}] \nonumber \\
& -4 \lambda_4  {\rm tr} \left[ B_{\{rv} B_{ri} \partial_v B_{a\}i} \right] -4 \lambda_5  {\rm tr} \left[ B_{rv} B_{ai} \partial_v B_{ai} + \frac{1}{2} B_{av} B_{(ri} \partial_v B_{a)i} \right] \nonumber \\
&-4\lambda_6 {\rm tr}(B_{av} B_{ai} \partial_v B_{ai}) - 4 \lambda_7 {\rm tr} \left[ B_{\{rk} B_{rl} \partial_k B_{a\}l} \right]
- 4 \lambda_8 {\rm tr} \left[ B_{\{rk} B_{al} \partial_k B_{a\}l} \right] \nonumber \\
&-4\lambda_9 {\rm tr}\left( B_{ak} B_{al} \partial_k B_{al} \right). \label{Leff_cubic}
\end{align}
Here, $r$-,$a$-indices inside $\{ \cdots \}$ and $(\cdots)$ shall be understood as all possible permutations, and symmetrization, respectively\footnote{Space-time indices $v,i$ inside $\{ \cdots \}$ and $(\cdots)$ are not affected.}, e.g.
\begin{align}
A_{\{rra\}} \equiv A_{rra} + A_{arr} + A_{rar}, \qquad \qquad A_{(ra)} = A_{ra} + A_{ar}.
\end{align}
The overall factor $4$ in \eqref{Leff_cubic} comes from the fact that $4\, {\rm tr}(t^{\rm a} t^{\rm b} t^{\rm c}) = i\epsilon^{\rm abc}$. Various coefficients in \eqref{Leff_cubic} are
\begin{align}
&\lambda_1 = i\log(2r_h) ,\qquad \qquad \, \lambda_2 =  \frac{ \pi}{48},\qquad \qquad \, \lambda_3 = \frac{i}{4}\log(2r_h), \nonumber \\
&\lambda_4 = -\frac{i}{2}\log(2r_h^2),\qquad \quad \lambda_5 = -\frac{\pi}{12},\qquad \quad \, \, \lambda_6 = -\frac{i\zeta(3)}{2 \pi^2}-\frac{i}{8}\log(2r_h^2), \nonumber \\
&\lambda_7 = i\log(r_h),\qquad \qquad \,\,\,\, \lambda_8 = \frac{\pi}{8},\qquad \qquad \,\,\, \, \lambda_9 = \frac{21i\zeta(3)}{16\pi^2}+\frac{i}{4}\log(r_h),  \label{lambda}
\end{align}
where $\zeta(x)$ is Riemann zeta function. By dimensional analysis, pieces such as $\log(r_h)$ etc. shall be understood as $\log(r_h/L)$ with $L$ the AdS radius which we set to unity.

~

$\bullet$ {\bf Derivative expansion of $\mathcal L_{eff}^{(4)}$}

~

The quartic effective Lagrangian $\mathcal L_{eff}^{(4)}$ starts from zeroth order in boundary derivative expansion:
\begin{align}
\mathcal L_{eff}^{(4)} = \mathcal L_{eff}^{(4),\; \rm LO} + \mathcal L_{eff}^{(4),\; \rm NLO} + \cdots.
\end{align}
For convenience of presentation, $\mathcal L_{eff}^{(4),\; \rm LO}$ may be split into four pieces:
\begin{align}
\mathcal L_{eff}^{(4),\; \rm LO} = \mathscr{L}_1 + \mathscr{L}_2 + \mathscr{L}_3 + \mathscr{L}_4,
\end{align}
where
\begin{align}
\mathscr L_1 = &\chi_{1,1} \left( B_{rv}^{\rm a} B_{rv}^{\rm a} B_{ri}^{\rm b} B_{ai}^{\rm b} + B_{rv}^{\rm a} B_{av}^{\rm a} B_{ri}^{\rm b} B_{ri}^{\rm b}\right) +\chi_{1,2}  \left( B_{rv}^{\rm a} B_{rv}^{\rm a} B_{ai}^{\rm b} B_{ai}^{\rm b} + B_{rv}^{\rm a} B_{av}^{\rm a} B_{ri}^{\rm b} B_{ai}^{\rm b} \right. \nonumber \\
& \left.+ B_{av}^{\rm a} B_{rv}^{\rm a} B_{ri}^{\rm b} B_{ai}^{\rm b}\right) +\chi_{1,3} \left( B_{rv}^{\rm a} B_{av}^{\rm a} B_{ai}^{\rm b} B_{ai}^{\rm b} + B_{av}^{\rm a} B_{av}^{\rm a} B_{ri}^{\rm b} B_{ai}^{\rm b} \right) +\chi_{1,4} B_{rv}^{\rm a} B_{av}^{\rm a} B_{ai}^{\rm b} B_{ai}^{\rm b} \nonumber \\
&+\chi_{1,5} B_{av}^{\rm a} B_{av}^{\rm a} B_{ai}^{\rm b} B_{ai}^{\rm b}, \nonumber \\
\mathscr L_2 = &\chi_{2,1} \left( B_{rv}^{\rm a} B_{rv}^{\rm b} B_{ri}^{\rm a} B_{ai}^{\rm b} + B_{rv}^{\rm a} B_{av}^{\rm b} B_{ri}^{\rm a} B_{ri}^{\rm b}\right) +\chi_{2,2} \left( B_{rv}^{\rm a} B_{rv}^{\rm b} B_{ai}^{\rm a} B_{ai}^{\rm b} + B_{rv}^{\rm a} B_{av}^{\rm b} B_{ri}^{\rm a} B_{ai}^{\rm b} \right. \nonumber \\
& \left.+ B_{av}^{\rm a} B_{rv}^{\rm b} B_{ri}^{\rm a} B_{ai}^{\rm b}\right) +\chi_{2,3}\left( B_{rv}^{\rm a} B_{av}^{\rm b} B_{ai}^{\rm a} B_{ai}^{\rm b} + B_{av}^{\rm a} B_{av}^{\rm b} B_{ri}^{\rm a} B_{ai}^{\rm b} \right) +\chi_{2,4} B_{rv}^{\rm a} B_{av}^{\rm b} B_{ai}^{\rm a} B_{ai}^{\rm b} \nonumber \\
&+\chi_{2,5} B_{av}^{\rm a} B_{av}^{\rm b} B_{ai}^{\rm a} B_{ai}^{\rm b}, \nonumber \\
\mathscr L_3 =& \chi_{3,1} B_{ri}^{\rm a} B_{ri}^{\rm a} B_{rj}^{\rm b} B_{aj}^{\rm b} +\chi_{3,2} \left( B_{ri}^{\rm a} B_{ri}^{\rm a} B_{aj}^{\rm b} B_{aj}^{\rm b} + B_{ri}^{\rm a} B_{ai}^{\rm a} B_{rj}^{\rm b} B_{aj}^{\rm b} + B_{ai}^{\rm a} B_{ri}^{\rm a} B_{rj}^{\rm b} B_{aj}^{\rm b} \right) \nonumber \\
&+\chi_{3,3}B_{ri}^{\rm a} B_{ai}^{\rm a} B_{aj}^{\rm b} B_{aj}^{\rm b} +\chi_{3,4}B_{ai}^{\rm a} B_{ai}^{\rm a} B_{aj}^{\rm b} B_{aj}^{\rm b}, \nonumber \\
\mathscr L_4 =&\chi_{4,1} B_{ri}^{\rm a} B_{ri}^{\rm b} B_{rj}^{\rm a} B_{aj}^{\rm b} +\chi_{4,2} \left( B_{ri}^{\rm a} B_{ri}^{\rm b} B_{aj}^{\rm a} B_{aj}^{\rm b} + B_{ri}^{\rm a} B_{ai}^{\rm b} B_{rj}^{\rm a} B_{aj}^{\rm b} + B_{ai}^{\rm a} B_{ri}^{\rm b} B_{rj}^{\rm a} B_{aj}^{\rm b} \right) \nonumber \\
&+\chi_{4,3}B_{ri}^{\rm a} B_{ai}^{\rm b} B_{aj}^{\rm a} B_{aj}^{\rm b} +\chi_{4,4}B_{ai}^{\rm a} B_{ai}^{\rm b} B_{aj}^{\rm a} B_{aj}^{\rm b}, \label{Leff_quartic_LO}
\end{align}
where various coefficients are
\begin{align}
&\chi_{1,1} = -\chi_{2,1} = -\log(2r_h),\qquad \chi_{1,2} = -\chi_{2,2} =\frac{i\pi}{48}, \qquad \chi_{1,3} = -\chi_{2,3} = - \frac{1}{4}\log(2r_h), \nonumber \\
&\chi_{1,4} = -\chi_{2,4} = - \frac{\zeta(3)}{16\pi^2},\qquad \quad \chi_{1,5} = -\chi_{2,5} =  \frac{i\pi}{192}, \qquad \chi_{3,1} = -\chi_{4,1} = \log(r_h), \nonumber \\
&\chi_{3,2} = -\chi_{4,2} = - \frac{i\pi}{16}, \qquad \chi_{3,3} = -\chi_{4,3} = \frac{1}{4}\log(r_h) + \frac{21\zeta(3)}{16\pi^2} ,\qquad \chi_{3,4} = -\chi_{4,4} = - \frac{i\pi}{128}. \label{chi}
\end{align}

Similarly, the next-to-leading order result $\mathcal L_{eff}^{(4),\; \rm NLO}$ is also split into four pieces
\begin{align}
\mathcal L_{eff}^{(4),\; \rm NLO}  = \tilde{\mathscr{L}}_1 + \tilde{\mathscr{L}}_2 + \tilde{\mathscr{L}}_3 + \tilde{\mathscr{L}}_4,
\end{align}
where
\begin{align}
\tilde{\mathscr{L}}_1 = &\tilde\chi_{1,1} \left( B_{rv}^{\rm a} B_{rv}^{\rm a} B_{ai}^{\rm b} \partial_v B_{ri}^{\rm b} - B_{rv}^{\rm a} B_{rv}^{\rm a} B_{ri}^{\rm b} \partial_v B_{ai}^{\rm b} + 2 B_{rv}^{\rm a} B_{av}^{\rm a} B_{ri}^{\rm b} \partial_v B_{ri}^{\rm b} \right) \nonumber \\
&+\tilde\chi_{1,2} \left( -B_{rv}^{\rm a} B_{rv}^{\rm a} B_{ai}^{\rm b} \partial_v B_{ai}^{\rm b} + 2 B_{rv}^{\rm a} B_{av}^{\rm a} B_{ai}^{\rm b} \partial_v B_{ri}^{\rm b} \right) +\tilde\chi_{1,3} \left( - \frac{1}{2} B_{rv}^{\rm a} B_{av}^{\rm a} B_{ai}^{\rm b} \partial_v B_{ai}^{\rm b} \right. \nonumber \\
&\left. + \frac{1}{4} B_{av}^{\rm a} B_{av}^{\rm a} B_{ai}^{\rm b} \partial_v B_{ri}^{\rm b} - \frac{1}{4} B_{av}^{\rm a} B_{av}^{\rm a} B_{ri}^{\rm b} \partial_v B_{ai}^{\rm b} \right) +\tilde\chi_{1,4}  B_{av}^{\rm a} B_{av}^{\rm a} B_{ai}^{\rm b} \partial_v B_{ai}^{\rm b},\nonumber \\
\tilde{\mathscr{L}}_2 = &\tilde\chi_{2,1} \left( B_{rv}^{\rm a} B_{rv}^{\rm b} B_{ai}^{\rm a} \partial_v B_{ri}^{\rm b} - B_{rv}^{\rm a} B_{rv}^{\rm b} B_{ri}^{\rm a} \partial_v B_{ai}^{\rm b} + B_{rv}^{\rm a} B_{av}^{\rm b} \partial_v B_{ri}^{\rm a} B_{ri}^{\rm b} + B_{rv}^{\rm a} B_{av}^{\rm b} B_{ri}^{\rm a} \partial_v B_{ri}^{\rm b}   \right) \nonumber \\
&+\tilde\chi_{2,2}  \left( -B_{rv}^{\rm a} B_{rv}^{\rm b} \partial_v B_{ai}^{\rm a} B_{ai}^{\rm b} + B_{rv}^{\rm a} B_{av}^{\rm b} \partial_v B_{ri}^{\rm a} B_{ai}^{\rm b} + B_{rv}^{\rm a} B_{av}^{\rm b} B_{ai}^{\rm a} \partial_v B_{ri}^{\rm b} \right) \nonumber \\
&+\tilde\chi_{2,3} \left( - \frac{1}{4} B_{rv}^{\rm a} B_{av}^{\rm b} \partial_v B_{ai}^{\rm a} B_{ai}^{\rm b} - \frac{1}{4} B_{rv}^{\rm a} B_{av}^{\rm b} B_{ai}^{\rm a} \partial_v B_{ai}^{\rm b} + \frac{1}{4} B_{av}^{\rm a} B_{av}^{\rm b} B_{ai}^{\rm a} \partial_v B_{ri}^{\rm b} \right.\nonumber \\
&\left. - \frac{1}{4} B_{av}^{\rm a} B_{av}^{\rm b} B_{ri}^{\rm a} \partial_v B_{ai}^{\rm b} \right) +\tilde\chi_{2,4}  B_{av}^{\rm a} B_{av}^{\rm b} \partial_v B_{ai}^{\rm a} B_{ai}^{\rm b},\nonumber \\
\tilde{\mathscr{L}}_3 = & \tilde\chi_{3,1} B_{ri}^{\rm a} B_{ri}^{\rm a} B_{rj}^{\rm b} \partial_v B_{aj}^{\rm b} +\tilde\chi_{3,2} \left( \partial_v B_{ri}^{\rm a} B_{ri}^{\rm a} B_{aj}^{\rm b} B_{aj}^{\rm b} + \partial_v B_{ri}^{\rm a} B_{ai}^{\rm a} B_{rj}^{\rm b} B_{aj}^{\rm b} + \partial_v B_{ri}^{\rm a} B_{ai}^{\rm a} B_{aj}^{\rm b} B_{rj}^{\rm b}  \right), \nonumber \\
\tilde{\mathscr{L}}_4 = & \tilde\chi_{4,1} B_{ri}^{\rm a} B_{ri}^{\rm b} B_{rj}^{\rm a} \partial_v B_{aj}^{\rm b} +\tilde\chi_{4,2} \left( \partial_v B_{ri}^{\rm a} B_{ri}^{\rm b} B_{aj}^{\rm a} B_{aj}^{\rm b} + \partial_v B_{ri}^{\rm a} B_{ai}^{\rm b} B_{aj}^{\rm a} B_{rj}^{\rm b} + B_{ai}^{\rm a} \partial_v B_{ri}^{\rm b} B_{aj}^{\rm a} B_{rj}^{\rm b}  \right),  \label{Leff_quartic_NLO}
\end{align}
where various coefficients are
\begin{align}
&\tilde\chi_{1,1} = -\tilde\chi_{2,1} = -\frac{\pi^2 \beta}{48},\qquad \tilde\chi_{1,2} = -\tilde\chi_{2,2} = - \frac{i\zeta(3) \beta}{16\pi}, \qquad \tilde\chi_{1,3} = -\tilde\chi_{2,3} =  - \frac{\pi^2 \beta}{48},\nonumber \\
&\tilde\chi_{1,4} = -\tilde\chi_{2,4} =   \frac{i\zeta(3) \beta}{32\pi}, \qquad \, \tilde\chi_{3,1} = -\tilde\chi_{4,1} =-\frac{\pi^2\beta}{8}, \qquad \quad  \tilde\chi_{3,2} = -\tilde\chi_{4,2} = \frac{i 21\zeta(3)\beta}{16\pi},  \label{chi_tilde}
\end{align}
where $\beta$ is the inverse temperature.

We compare our results with recent works on hydrodynamic EFT for charge diffusion, see e.g. \cite{Crossley:2015evo,Jensen:2017kzi,Jensen:2018hhx,Chen-Lin:2018kfl,Jain:2020zhu,Glorioso:2020loc}. The main novelty of present work could be identified as the following two aspects:

First, our effective action is more complete in systematically capturing nonlinear interactions among noise variable $\varphi_a^{\rm a}$, as well as nonlinear interactions between noise variable $\varphi_a^{\rm a}$ and dynamical variable $\varphi_r^{\rm a}$. In our effective action $S_{eff}$, these are represented by terms cubic and/or quartic in $B_{a\mu}^{\rm a}$, which are not considered in \cite{Glorioso:2020loc} (see also \cite{Jensen:2017kzi,Jensen:2018hhx,Chen-Lin:2018kfl} for U(1) diffusion). As pointed out in \cite{Crossley:2015evo}, nonlinear interactions of this type cannot be covered in stochastic formulation of hydrodynamics. Therefore, it will be interesting to explore non-negligible signatures, generated by these nonlinear interactions, in hydrodynamic limit of correlators, following the example of U(1) diffusion \cite{Jain:2020zhu}.

Second, derived within a holographic model, our effective action $S_{eff}$ automatically accounts for both thermal fluctuation and quantum fluctuation, in contrast with \cite{Crossley:2015evo,Chen-Lin:2018kfl,Jain:2020zhu,Glorioso:2020loc} which focused on thermal noise. A quantum hydrodynamic EFT for U(1) diffusion was considered in \cite{Jensen:2017kzi,Jensen:2018hhx}, which implemented the dynamical KMS symmetry at quantum level. It would be interesting to clarify consequences of quantum fluctuations based on quantum hydrodynamic EFT as constructed here by us and in \cite{Jensen:2017kzi,Jensen:2018hhx}.

Finally, our results perfectly pass through {\bf various consistency checks}:

$\bullet$ $Z_2$-reflection symmetry

Basically, this requires the effective action $S_{eff}$ to satisfy
\begin{align}
&S_{eff}^* [B_{1\mu}^{\rm a}(x), B_{2\mu}^{\rm a}(x)] = - S_{eff}[ B_{2\mu}^{\rm a}(x), B_{1\mu}^{\rm a}(x)] \nonumber \\
\Leftrightarrow & S_{eff}^* [B_{r\mu}^{\rm a}(x), B_{a\mu}^{\rm a}(x)] = - S_{eff}[ B_{r\mu}^{\rm a}(x), - B_{a\mu}^{\rm a}(x)], \label{Z2_reflection}
\end{align}
which implies that the effective Lagrangian must have some complex coefficients. When field contents are real variables (as in our case), the $Z_2$ reflection symmetry \eqref{Z2_reflection} requires that coefficient of a term containing even number of $a$-type variables must be purely imaginary, while coefficient of a term containing odd number of $a$-type variables must be purely real. This is perfectly satisfied by our results.

$\bullet$ Imaginary part of $S_{eff}$ is non-negative.

This requirement is to ensure that path integral based on effective action $S_{eff}$ is well-defined. For quadratic Lagrangian $\mathcal L_{eff}^{(2)}$, this is perfectly satisfied: leading order results of $w_1, w_2$ are non-negative, see \eqref{w_expansion}. For cubic Lagrangian $\mathcal L_{eff}^{(3)}$, this requirement does not give any constraint. At quartic order, we collect positive-definite structures
\begin{align}
\mathcal L_{\rm pd}^{(4)} &= \chi_{1,2} B_{rv}^{\rm a} B_{rv}^{\rm a} B_{ai}^{\rm b} B_{ai}^{\rm b} + \chi_{2,2} B_{rv}^{\rm a} B_{rv}^{\rm b} B_{ai}^{\rm a} B_{ai}^{\rm b} + \chi_{1,5} B_{av}^{\rm a} B_{av}^{\rm a} B_{ai}^{\rm b} B_{ai}^{\rm b} \nonumber \\
&+ \chi_{2,5} B_{av}^{\rm a} B_{av}^{\rm b} B_{ai}^{\rm a} B_{ai}^{\rm b}
+ \chi_{3,2} B_{ri}^{\rm a} B_{ri}^{\rm a} B_{aj}^{\rm b} B_{aj}^{\rm b} + \chi_{4,2} B_{ri}^{\rm a} B_{ri}^{\rm b} B_{aj}^{\rm a} B_{aj}^{\rm b} \nonumber \\
&+ \chi_{3,4} B_{ai}^{\rm a} B_{ai}^{\rm a} B_{ai}^{\rm b} B_{ai}^{\rm b} + \chi_{4,4} B_{ai}^{\rm a} B_{ai}^{\rm b} B_{aj}^{\rm a} B_{aj}^{\rm b}, \label{L_positive_definite}
\end{align}
where each term (apart of the coefficient) is non-negative for any configuration of $B_{r\mu}^{\rm a}, B_{a\mu}^{\rm a}$. However, our results for the $\chi$'s in \eqref{L_positive_definite} are not all positive-definite. It is important to stress that this does not necessarily mean our results are pathological. Recall that our procedure of constructing $S_{eff}$ is based on perturbative expansion around origin of field configuration. Thus, our results are meaningful when $B_{r\mu}^{\rm a}, B_{a\mu}^{\rm a} $ are tiny, in which region we always have
\begin{align}
\frac{1}{2}w_1^0 B_{av}^{\rm a} B_{av}^{\rm a} + \frac{1}{2} w_2^0 B_{ai}^{\rm a} B_{ai}^{\rm a} \gg \big|\mathcal L_{\rm pd}^{(4)} \big|,
\end{align}
which guarantees the path integral to be well-defined. Here, $w_1^0, w_2^0$ are leading order results for $w_1,w_2$, see \eqref{w_expansion}. Not surprisingly, the fact that $\mathcal L_{\rm pd}^{(4)}$ is not positive-definite for ``big'' configuration of $B_{r\mu}^{\rm a}, B_{a\mu}^{\rm a}$ implies instabilities of the system, which we believe are related to those revealed in \cite{Gubser:2008wv,Ammon:2011je,Bu:2012mq}. It is of great interest to explore the exact relationship, and particularly construct an EFT for the order parameter associated with phase transitions investigated in \cite{Gubser:2008wv,Ammon:2011je,Bu:2012mq}, along the line of \cite{Bu:2021clf}. To this end, dynamical variable in the EFT will be identified with ``normalizable modes'' of bulk field $C_\mu^{\rm a}$, rather than non-normalizable modes employed here. We leave further investigation on this subject for future work.

$\bullet$ Time-independent diagonal gauge transformation of $B_{s\mu}$:
\begin{align}
&B_{1v} \to \mathcal U(\vec x) B_{1v} \mathcal U^{\dagger}(\vec x), \qquad \qquad \qquad \qquad \quad B_{2v} \to \mathcal U(\vec x) B_{2v} \mathcal U^{\dagger}(\vec x), \nonumber \\
&B_{1i} \to \mathcal U(\vec x) B_{1i} \mathcal U^{\dagger}(\vec x) + i\, \mathcal U(\vec x) \partial_i \mathcal U^{\dagger}(\vec x), \qquad B_{2i} \to \mathcal U(\vec x) B_{2i} \mathcal U^{\dagger}(\vec x) + i\, \mathcal U(\vec x) \partial_i \mathcal U^{\dagger}(\vec x),
\end{align}
where $\mathcal U(\vec x)= e^{i \alpha^{\rm a}(\vec x) t^{\rm a}}$ is an element of the SU(2) group for the boundary theory. Equivalently, the above symmetry requirement is
\begin{align}
&B_{rv} \to \mathcal U(\vec x) B_{rv} \mathcal U^{\dagger}(\vec x), \qquad B_{a\mu} \to \mathcal U(\vec x) B_{a\mu} \mathcal U^{\dagger}(\vec x), \nonumber \\
&B_{ri} \to \mathcal U(\vec x) B_{ri} \mathcal U^{\dagger}(\vec x) + i \mathcal U(\vec x) \partial_i \mathcal U^{\dagger}(\vec x), \label{gauge_symmetry_diagonal}
\end{align}
which relates certain coefficients in $\mathcal L_{eff}^{(3)}$ and $\mathcal L_{eff}^{(4),\, \rm LO}$:
\begin{align}
&\lambda_1  = -i\chi_{1,1} = i\chi_{2,1}, \qquad  \lambda_2  = -i\chi_{1,2} = i\chi_{2,2}, \qquad \quad \lambda_3  = -i\chi_{1,3} = i\chi_{2,3}\nonumber \\
&\lambda_7  = i\chi_{3,1} = -i\chi_{4,1}, \qquad  \lambda_8  = 2i\chi_{3,2} = -2i\chi_{4,2}, \qquad  \lambda_9  = i\chi_{3,3} = -i\chi_{4,3},
\end{align}
which are perfectly obeyed by holographic results \eqref{lambda} and \eqref{chi}.

Indeed, with the help of covariant derivative operator $\mathcal{D}_i \equiv \partial_i - i [B_{ri}, ~~] $ as invented in \cite{Glorioso:2020loc}, our results could be reorganized into a more compact form\footnote{We have not considered putting $\mathcal L_{eff}^{(4),\, \rm NLO}$ into a form bearing symmetry \eqref{gauge_symmetry_diagonal} transparently, which would require some second and third order derivative terms that are beyond scope of present work.}:
\begin{align}
	\mathcal L_{eff}^{(2)} + \mathcal L_{eff}^{(3)} + \mathcal L_{eff}^{(4),\, \rm LO} = \tilde{\mathcal L}_{eff}^{(2)} + \tilde{\mathcal L}_{eff}^{(3)} + \tilde{\mathcal L}_{eff}^{(4), \, \rm LO}, \label{Leff_reorganize}
\end{align}
where
\begin{align}
\tilde{\mathcal L}_{eff}^{(2)} = & \frac{i}{2} B_{av}^{\rm a} w_{1} B_{av}^{\rm a} + \frac{i}{2} B_{ai}^{\rm a} w_{2} B_{ai}^{\rm a} + \frac{i}{2} \mathcal{D}_{i} B_{ai}^{\rm a} w_{3} \mathcal{D}_{j} B_{aj}^{\rm a} + i B_{av}^{\rm a} w_{4} \mathcal{D}_{k} B_{ai}^{\rm a} + B_{av}^{\rm a} w_{5} B_{rv}^{\rm a} \nonumber \\
& + B_{av}^{\rm a} w_{6} \mathcal{D}_{i} \partial_{v} B_{ri}^{\rm a} + \mathcal{D}_{i} B_{ai}^{\rm a} w_{7} B_{rv}^{\rm a} + B_{ai}^{\rm a} w_{8} \partial_{v} B_{ri}^{\rm a} + \mathcal{D}_i B_{aj}^{\rm a} w_{9} \mathcal{F}_{rij}^{\rm a}, \label{Leff_quadratic_cov}\\
\tilde{\mathcal L}_{eff}^{(3)} = & -i \lambda_{1} \epsilon^{\rm abc} B_{rv}^{\rm a} B_{ai}^{\rm b} \mathcal{D}_{i} B_{rv}^{\rm c} -i \lambda_{2} \epsilon^{\rm abc} (B_{rv}^{\rm a} B_{ai}^{\rm b} \mathcal{D}_{i} B_{av}^{\rm c} + B_{av}^{\rm a} B_{ai}^{\rm b} \mathcal{D}_{i} B_{rv}^{\rm c})\nonumber \\
& -i \lambda_{3} \epsilon^{\rm abc} B_{av}^{\rm a} B_{ai}^{\rm b} \mathcal{D}_{i} B_{av}^{\rm c} - 2 i \lambda_{4} \epsilon^{\rm abc} B_{rv}^{\rm a} B_{ai}^{\rm b} \partial_{v} B_{ri}^{\rm c} - i \lambda_{5} \epsilon^{\rm abc} (B_{rv}^{\rm a} B_{ai}^{\rm b} \partial_{v} B_{ai}^{\rm c} \nonumber \\
&+ B_{av}^{\rm a} B_{ai}^{\rm b} \partial_{v} B_{ri}^{\rm c}) -i \lambda_{6} \epsilon^{\rm abc} B_{av}^{\rm a} B_{ai}^{\rm b} \partial_{v} B_{ai}^{\rm c} -i \lambda_{8} \epsilon^{\rm abc} B_{ai}^{\rm a} B_{aj}^{\rm b} \mathcal{F}_{rij}^{\rm c}\nonumber \\
&-i \lambda_{9} \epsilon^{\rm abc} B_{ai}^{\rm a} B_{aj}^{\rm b} \mathcal{D}_{i} B_{aj}^{\rm c}, \label{Leff_cubic_cov}\\
\tilde{\mathcal L}_{eff}^{(4), \, \rm LO} = & \chi_{1,2} B_{rv}^{\rm a} B_{rv}^{\rm a} B_{ai}^{\rm b} B_{ai}^{\rm b}+ (\chi_{1,3} + \chi_{1,4}) B_{rv}^{\rm a} B_{av}^{\rm a} B_{ai}^{\rm b} B_{ai}^{\rm b} +  \chi_{1,5} B_{av}^{\rm a} B_{av}^{\rm a} B_{ai}^{\rm b} B_{ai}^{\rm b} \nonumber \\
&+\chi_{2,2} B_{rv}^{\rm a} B_{rv}^{\rm b} B_{ai}^{\rm a} B_{ai}^{\rm b} + (\chi_{2,3} + \chi_{2,4}) B_{rv}^{\rm a} B_{av}^{\rm b} B_{ai}^{\rm a} B_{ai}^{\rm b} +  \chi_{2,5} B_{av}^{\rm a} B_{av}^{\rm b} B_{ai}^{\rm a} B_{ai}^{\rm b}\nonumber  \\
& + \chi_{3,4}  B_{ai}^{\rm a} B_{ai}^{\rm a} B_{aj}^{\rm b} B_{aj}^{\rm b} + \chi_{4,4}  B_{ai}^{\rm a} B_{ai}^{\rm b} B_{aj}^{\rm a} B_{aj}^{\rm b}, \label{Leff_quartic_cov}
\end{align}
where all the coefficients are constants except for the following ones in $\tilde{\mathcal L}_{eff}^{(2)}$
\begin{align}
& w_m = w_m^0 + w_m^1 \partial_v + w_m^{2,0} \partial_v^2 +  w_m^{0,2} \mathcal{D}^2, \qquad m= 1,2,5 \nonumber \\
& w_n = w_n^0 + w_n^1 \partial_v, \qquad n = 4,7,8.
\end{align}
In order to achieve \eqref{Leff_quadratic_cov}, we have utilized those results for second order derivative terms for quadratic action of \cite{Bu:2020jfo}, namely, the values for $w_m^{2,0}, w_m^{0,2}$ ($m=1,2,5$), $w_n^1$ ($n=4,7,8$), and $w_9$. As already noticed in \cite{Glorioso:2020loc}, the ``building blocks'', say $B_{a\mu}$, $\mathcal{D}_i B_{a\mu}$, $B_{rv}$, $\mathcal{D}_i B_{rv}$, $\partial_v B_{ri}$, $\mathcal F_{rij} \equiv \partial_i B_{rj} - \partial_j B_{ri} - i[B_{ri}, B_{rj}]$, transform in a simple way under \eqref{gauge_symmetry_diagonal}, schematically $(\cdots) \to \mathcal{U}(\vec x) (\cdots) \mathcal U^\dagger(\vec x)$. Thus, the symmetry \eqref{gauge_symmetry_diagonal} becomes more manifest for \eqref{Leff_reorganize}.

$\bullet$ Dynamical KMS symmetry at quantum level.

The proposal of dynamical KMS symmetry \cite{Glorioso:2017fpd} is crucial in formulating non-equilibrium EFT for dissipative fluids. While it is usually implemented in the classical statistical limit (i.e., capturing only thermal fluctuations), we find that our effective action $S_{eff}$ derived from holographic model satisfies dynamical KMS symmetry at quantum level, which implies the boundary system, although in the large $N_c$ limit, is actually a quantum system as expected. We defer more details to subsection \ref{holo_KMS_symmetry}.

$\bullet$ Current constitutive relation in the mean-field limit.

With the effective action $S_{eff}$, we compute the hydrodynamic constitutive relation by taking mean-field limit \cite{Crossley:2015evo,Glorioso:2020loc}:
\begin{align}
J^{{\rm a}\,\mu}_{\rm mf} \equiv \frac{\delta S_{eff}}{\delta A_\mu^{\rm a}} \bigg|_{B_{a\mu} =0}.
\end{align}
When the SU(2) background electromagnetic field is turned off, the spatial component of the hydrodynamic current is
\begin{align}
J^{{\rm a}\, i }_{\rm mf} = w_8^0 \partial_i \mu^{\rm a} + i(2\lambda_{4}-\lambda_{1}) \epsilon^{\rm abc} \mu^{\rm b} \partial_i \mu^{\rm c} + \tilde \chi_{2,1} \left( \mu^a \mu^{\rm b} \partial_i \mu^{\rm b} - \mu^{\rm b} \mu^{\rm b} \partial_i \mu^{\rm a} \right), \label{Jmu_mf}
\end{align}
where the SU(2) chemical potential is defined as \cite{Glorioso:2020loc}
\begin{align}
\mu \equiv \mathcal U^{-1}(\varphi) B_{rv} \mathcal U(\varphi) = A_v - i \mathcal U^{-1}(\varphi) \partial_v \mathcal U(\varphi)
\end{align}
Before comparing \eqref{Jmu_mf} with relevant results of \cite{Torabian:2009qk,Glorioso:2020loc}, we clarify similarities/differences between the models studied in \cite{Torabian:2009qk,Glorioso:2020loc} and the one explored here. Apart of backreaction effect, our holographic model is the same as that of \cite{Torabian:2009qk}. While non-Abelian SU(2) symmetry is essential in both \cite{Torabian:2009qk,Glorioso:2020loc} and present work, the conserved non-Abelian charges are identified as spin densities in \cite{Glorioso:2020loc}, which is different from the flavor charges in \cite{Torabian:2009qk} and this work. Now, we turn to compare our result with \cite{Torabian:2009qk,Glorioso:2020loc}. Our result \eqref{Jmu_mf} slightly differs from that of \cite{Glorioso:2020loc}, particularly on the structure of the last term. However, \eqref{Jmu_mf} is fully consistent with \cite{Torabian:2009qk}, which derived the current constitutive relation by fluid-gravity correspondence.

\subsection{Dynamical KMS symmetry: from boundary to bulk} \label{holo_KMS_symmetry}

With the effective action $S_{eff}[B_r, B_a]$, the partition function of dual boundary theory is represented as a path integral over gapless modes (for notational simplification we omitted SU(2) flavor indices):
\begin{align}
Z_{\rm CFT} = \int [D \varphi_r] [D \varphi_a] e^{i S_{eff}[B_{r\mu}, B_{a \mu}]}. \label{Z_CFT1}
\end{align}
The doubling of degrees of freedom, due to usage of SK formalism, guarantees systematic inclusion of both fluctuations and dissipations in the boundary theory. The information of state is reflected in coefficients of effective action $S_{eff}$. It turns out that a path integral like \eqref{Z_CFT1} actually corresponds to quantum field theory of a statistical system, in which both statistical fluctuations and quantum fluctuations are consistently covered.

When the boundary system is in a thermal state, the KMS condition sets important constraint on the generating functional $W = -i \log Z_{\rm CFT}$. The KMS condition can be expressed in terms of $n$-point correlation functions (i.e., functional derivatives of $W$ with respect to external sources $A_{s\mu}$), generalizing familiar FDT to nonlinear case \cite{Wang:1998wg,Hou:1998yc} (see also \cite{Crossley:2015evo}). Obviously, the KMS condition and the generalized nonlinear FDT are valid at the full quantum level. Within non-equilibrium EFT framework, KMS condition is guaranteed by the proposal that non-equilibrium effective action $S_{eff}$ shall satisfy dynamical KMS symmetry \cite{Glorioso:2016gsa,Glorioso:2017fpd}:
\begin{align}
S_{eff}[B_{1\mu}, B_{2\mu}] = S_{eff}[\widehat B_{1\mu}, \widehat B_{2\mu}],  \label{dynamical_KMS_quantum_action}
\end{align}
where
\begin{align}
\widehat B_{1\mu}(-v,-\vec x) = (-1)^{\eta_\mu}B_{1\mu}(v, \vec x), \qquad \qquad \widehat B_{2\mu}(-v,-\vec x) = (-1)^{\eta_\mu}B_{2\mu}(v- i \beta, \vec x), \label{dynamical_KMS_quantum}
\end{align}
where $\beta$ is the inverse temperature, and $(-1)^{\eta_\mu}$ is the eigenvalue of discrete symmetry transformation $\Theta$ (containing time-reversal $\mathcal T$) acting on $B_\mu$. Physically, the dynamical KMS symmetry \eqref{dynamical_KMS_quantum} plays the role of imposing microscopic time-reversibility and local equilibrium \cite{Glorioso:2016gsa}. Interestingly, it was discovered that by taking classical statistical limit \cite{Glorioso:2016gsa,Glorioso:2017fpd}, the theory \eqref{Z_CFT1} can be consistently truncated into a classical statistical theory, in which only statistical thermal fluctuations survive. To this end, one restores the Planck constant $\hbar$ by substituting $\beta \to \hbar\beta$ in \eqref{dynamical_KMS_quantum} and writes
\begin{align}
B_{1\mu} = B_{r\mu} + \frac{\hbar}{2} B_{a\mu}, \qquad B_{2\mu} = B_{r\mu} - \frac{\hbar}{2} B_{a\mu}.
\end{align}
Then, taking the limit $\hbar \to 0$, one obtains classical statistical limit of the dynamical KMS symmetry \eqref{dynamical_KMS_quantum}
\begin{align}
\widehat B_{r\mu}(-v,-\vec x) = (-1)^{\eta_\mu}B_{r\mu}(v, \vec x), \quad \widehat B_{a\mu}(-v,-\vec x) = (-1)^{\eta_\mu} \left[ B_{a\mu}(v, \vec x) + i \beta \partial_0 B_{r\mu}(v, \vec x) \right]. \label{dynamical_KMS_classical}
\end{align}
In the comprehensive studies of \cite{Crossley:2015evo,Glorioso:2016gsa,Glorioso:2017fpd}, it is indeed the classical statistical limit \eqref{dynamical_KMS_classical} that was implemented for the construction of hydrodynamic EFT for dissipative charged fluids. Thus, the effective theory constructed in \cite{Crossley:2015evo,Glorioso:2016gsa,Glorioso:2017fpd} covers statistical thermal fluctuations but ignores quantum ones. Later on, this was refined in \cite{Blake:2017ris} which proposed a quantum hydrodynamic theory (valid at finite $\hbar$ and to all orders in derivatives) for maximally chaotic systems (see also \cite{Jensen:2017kzi,Jensen:2018hhx} for the problem of U(1) diffusion).

Now we turn to the effective action $S_{eff}$ derived within a specific holographic model. While our derivation is carried out in the large $N_c$ limit such that the dual gravity becomes classical, this does not necessarily mean the dual boundary theory will only capture thermal fluctuations. Thus, we do not expect our effective action $S_{eff}$ to obey the classical statistical limit \eqref{dynamical_KMS_classical}. Moreover, the derivative expansion adopted in the holographic derivation corresponds to $\beta$-expansion on the boundary theory. Thus, instead of the $\hbar$-expansion, it is reasonable to consider $\beta$-expansion of \eqref{dynamical_KMS_quantum}
\begin{align}
&\widehat B_{r\mu}(-v,- \vec x) = (-1)^{\eta_\mu} \left[B_{r\mu}(v,\vec x) - \frac{i}{2} \hbar \beta \partial_0 B_{r\mu}(v,\vec x) + \frac{i}{4} \hbar^2\beta \partial_0 B_{a\mu}(v,\vec x) \right], \nonumber \\
&\widehat B_{a\mu}(-v,- \vec x) = (-1)^{\eta_\mu} \left[B_{a\mu}(v,\vec x) + i \beta \partial_0 B_{r\mu}(v,\vec x) - \frac{i}{2} \hbar \beta \partial_0 B_{a\mu}(v,\vec x) \right], \label{dynamical_KMS_beta_expansion}
\end{align}
which reduces into the classical statistical limit \eqref{dynamical_KMS_classical} once $\hbar \to 0$ is taken. Given that our $S_{eff}$ is valid to first order in boundary derivative, in \eqref{dynamical_KMS_beta_expansion} we ignored higher powers in $\beta$. The dynamical KMS symmetry \eqref{dynamical_KMS_beta_expansion} puts constraints\footnote{For quadratic Lagrangian $\mathcal L_{eff}^{(2)}$, dynamical KMS symmetry has been carefully examined in \cite{Bu:2020jfo} beyond hydrodynamic limit.} among some coefficients in $\mathcal L_{eff}^{(4),\, \rm LO}$ and $\mathcal L_{eff}^{(4),\, \rm NLO}$:
\begin{align}
& \tilde\chi_{1,1} = -\tilde\chi_{2,1}=\frac{i\beta}{2}\chi_{1,2}, \qquad \tilde\chi_{1,2} = -\tilde\chi_{2,2}=\frac{i\beta}{2} \left(\chi_{1,3} +\chi_{1,4} -\frac{1}{4}\chi_{1,1} \right), \nonumber \\
& \tilde\chi_{1,3} = -\tilde\chi_{2,3}=\frac{i\beta}{2}\chi_{1,2}, \qquad \tilde\chi_{1,4} = -\tilde\chi_{2,4}=-\frac{i\beta}{4}\chi_{1,4}, \qquad \tilde\chi_{3,1} = -\tilde\chi_{4,1} = - i\beta \chi_{3,2}, \nonumber \\
& \tilde\chi_{3,2} = -\tilde\chi_{4,2}=\frac{i\beta}{2} \left( \chi_{3,3} - \frac{1}{4} \chi_{3,1} \right).  \label{KMS_chi_chi_tilde}
\end{align}
It is then straightforward to check that our holographic results \eqref{chi} and \eqref{chi_tilde} perfectly satisfy \eqref{KMS_chi_chi_tilde}\footnote{
While KMS-invariance of $S_{eff}^{(2)}$ and $S_{eff}^{(4)}$ is insensitive to the eigenvalue $(-1)^{\eta_\mu}$, for the cubic part $S_{eff}^{(3)}$ to be KMS-invariant, we shall think of the eigenvalue $(-1)^{\eta_\mu}$ to be associated with $\Theta=\mathcal{PT}$, say $(-1)^{\eta_\mu} =+1$ for all components, which shall be compensated by transpose of matrix-valued generators $t^{\rm a}$.}. This implies that effective theory derived from holographic method does capture both quantum and thermal fluctuations, as speculated in \cite{deBoer:2018qqm}. Moreover, in contrast with the hydrodynamic EFT framework, it is impossible to split quantum and thermal fluctuations for a holographic theory, as seen from \eqref{dynamical_KMS_beta_expansion}. More precisely, it is natural to think of $\hbar$-expansion and derivative expansion (i.e., $\beta$-expansion) as independent from hydrodynamic EFT perspective, the derivative expansion is controlled by the combination $\hbar \beta$ for a holographic theory.

Finally, we would like to understand implication of dynamical KMS symmetry \eqref{dynamical_KMS_quantum} on the bulk dynamics, and hopefully give a holographic interpretation of \eqref{dynamical_KMS_quantum}. Recall that we work in the saddle-point approximation for the bulk theory. Thus, we are motivated to examine properties of partially on-shell bulk solution under the KMS transformation \eqref{dynamical_KMS_quantum}. Via time-translational operator, we may rewrite the KMS transformation \eqref{dynamical_KMS_quantum} as \cite{Blake:2017ris}
\begin{align}
\widehat B_{1\mu}(-v,-\vec x)=(-1)^{\eta_\mu} B_{1\mu}(v,\vec x), \qquad \qquad \widehat B_{2\mu}(-v,-\vec x)=(-1)^{\eta_\mu} e^{-i\beta \partial_v} B_{2\mu}(v,\vec x). \label{KMS_transform_B12}
\end{align}
Then, it is direct to check that when boundary theory undergoes KMS transformation \eqref{KMS_transform_B12}, the leading order bulk solutions \eqref{Cperp_solution_LO} and \eqref{Cparallel_solution_LO} transform analogously
\begin{align}
&\widehat C_M^{{\rm a}(1)~ \rm up}(r,-x^\mu) = (-1)^{\eta_M} e^{-i \beta \partial_v} C_M^{{\rm a}(1)~ \rm up}(r,x^\mu), \nonumber \\
&\widehat C_M^{{\rm a}(1)~ \rm dw}(r,-x^\mu) = (-1)^{\eta_M} C_M^{{\rm a}(1)~ \rm dw}(r,x^\mu). \label{KMS_bulk_LO}
\end{align}
Here, the $\mathcal{PT}$-symmetries of the leading order solutions \eqref{Cperp_solution_LO} and \eqref{Cparallel_solution_LO}, extensively explored in \cite{Bu:2020jfo}, are useful in demonstrating the transformation property \eqref{KMS_bulk_LO}. It is important to stress that \eqref{KMS_bulk_LO} is valid to all orders in boundary derivatives. Indeed, \eqref{KMS_bulk_LO} amounts to saying that leading order EOMs \eqref{dynamical_EOM_1st} are invariant under the bulk KMS transformation \eqref{KMS_bulk_LO}, which is more transparent as viewed in Schwarzschild coordinate system.

We turn to higher order solution $C_M^{{\rm a}(n\geq2)}$, whose dynamical EOMs differ from those of $C_M^{{\rm a}(1)}$ by the source terms, $\mathbb S_\mu^{{\rm a}(1)}=0$ versus $\mathbb S_\mu^{{\rm a}(n\geq2)} \neq 0$. Iteratively, one can show that under KMS transformation \eqref{KMS_transform_B12}, the source term $\mathbb S_\mu^{{\rm a}(n\geq2)}$ changes analogously as \eqref{KMS_bulk_LO}
\begin{align}
&\widehat{\mathbb S}_\mu^{{\rm a}(n)\;\rm up}(r,-x^\alpha) = (-1)^{\eta_\mu} e^{-i\beta \partial_v}\mathbb S_\mu^{{\rm a}(n)\;\rm up}(r,x^\alpha), \nonumber \\
&\widehat{\mathbb S}_\mu^{{\rm a}(n)\;\rm dw}(r,-x^\alpha) = (-1)^{\eta_\mu} \mathbb S_\mu^{{\rm a}(n)\;\rm dw}(r,x^\alpha), \qquad \qquad  n=2,3,\cdots,
\end{align}
which, combined with vanishing AdS boundary conditions for $C_\mu^{{\rm a}(n\geq2)}$, helps to conclude that under the transformation \eqref{KMS_transform_B12}, the higher order bulk solution $C_M^{{\rm a}(n\geq2)}$ changes as
\begin{align}
&\widehat C_M^{{\rm a}(n)\; \rm up}(r,-x^\mu) = (-1)^{\eta_M} e^{-i \beta \partial_v} C_M^{{\rm a}(n)\; \rm up}(r,x^\mu), \nonumber \\
&\widehat C_M^{{\rm a}(n)\; \rm dw}(r,-x^\mu) = (-1)^{\eta_M} C_M^{{\rm a}(n)\; \rm dw}(r,x^\mu). \label{KMS_bulk_higher_order}
\end{align}

Eventually, we observe a holographic analogue of the dynamical KMS symmetry \eqref{dynamical_KMS_quantum} satisfied by the boundary effective theory \eqref{dynamical_KMS_quantum_action}: the bulk action $S_{\rm bulk} = S_0 + S_{\rm ct}$ shall satisfy the following genuine $Z_2$ symmetry:
\begin{align}
S_{\rm bulk}[C_M^{\rm up}, C_M^{\rm dw}] = S_{\rm bulk}[\widehat C_M^{\rm up}, \widehat C_M^{\rm dw}], \label{holo_KMS_action}
\end{align}
where
\begin{align}
\widehat C_M^{\rm up}(r,-x^\mu) = (-1)^{\eta_M} e^{-i \beta \partial_v} C_M^{\rm up}(r,x^\mu), \qquad \widehat C_M^{\rm dw}(r,-x^\mu) = (-1)^{\eta_M} C_M^{\rm dw} (r,x^\mu), \label{holo_KMS_symmetry_transform}
\end{align}
Thus, \eqref{holo_KMS_action} and \eqref{holo_KMS_symmetry_transform} are taken as holographic interpretation of dynamical KMS symmetry proposal \eqref{dynamical_KMS_quantum_action} and \eqref{dynamical_KMS_quantum}. The invariance of bulk action under bulk KMS transformation becomes more transparent if one turns to Fourier space, so that the exponential factor becomes unity due to delta-function of four-momenta.

\section{Summary and discussion} \label{summary}

In this work we present a holographic derivation of hydrodynamic EFT for SU(2) diffusion. To first order in derivative expansion, the effective action is analytically computed to quartic order in the gauge-invariant objects $B_{r\mu}^{\rm a}$ and $B_{a\mu}^{\rm a}$. As a non-Gaussian effective theory, such an EFT contains a complete list of cubic and quartic interactions. Particularly, the generalized nonlinear FDT is guaranteed via dynamical KMS symmetry at quantum level. The dynamical KMS symmetry is found to have a bulk analogue.

The effective action $S_{eff}$ provides a framework to systematically explore phenomenological consequences of nonlinear interactions as emphasized in subsection \ref{boundary_action}. One approach towards this end would be to compute loop diagrams based on the effective action, as in \cite{Chen-Lin:2018kfl,Jain:2020zhu,Jain:2020hcu,Sogabe:2021wqk,Sogabe:2021svv}. An alternative way would be to cast the hydrodynamic effective action into a stochastic differential equation satisfied by dynamical variable $\varphi_r$. However, as demonstrated in \cite{Crossley:2015evo}, the latter approach works perfectly only when terms beyond quadratic order in $\varphi_a$ are ignored in the effective action. Physically, this truncation amounts to turning off non-Gaussian noise, which is usually unavoidable and might be of importance in realistic systems \cite{Crossley:2015evo,Jain:2020zhu}.

It would be interesting to derive Fokker-Planck (FP) type equation for distribution of the dynamical variable $\varphi_r$, which is more suitable for numerical investigation. In contrast to stochastic Langevin-type equation, the FP equation describes the probability of finding the system in a certain configuration and is thus fully deterministic. When first order derivatives in $S_{eff}^{(3)}$ and $S_{eff}^{(4)}$ are turned off, this derivation can closely follow the textbook \cite{Kamenev2011} by making an analog of the hydrodynamic EFT with Hamiltonian formulation of quantum mechanics. By this analog, the task boils down to looking for ``Hamiltonian'' $H$ of the hydrodynamic EFT. A more general method would be to discretize the time and space, and consider the ``restricted'' partition function \cite{Kamenev2011}, which is identified as the probability of finding the system in a certain configuration. We leave such a derivation for a future project.

Finally, based on present work it is possible to go beyond classical treatment for the bulk theory, and consider loop corrections to tree-level Witten diagrams of Figure \ref{witten_diagram}. This corresponds to including finite $N_c$ correction in the boundary EFT \cite{Kovtun:2003vj,Caron-Huot:2009kyg}. Technically, this task will boil down to assembling bulk-to-boundary propagator (to be read off from leading order solution $C_\mu^{{\rm a}(1)}$) and bulk-to-bulk propagator (encoded in bulk Green's functions $G_{\perp, \parallel}$) using proper vertices. We leave this study as a forthcoming project.

\appendix

\section{Yang-Mills equation in different coordinate systems} \label{YM_eom_explicit}

In this appendix, we collect explicit forms of bulk Yang-Mills equation in two coordinate systems: the Schwarzschild versus ingoing EF.

From Schwarzschild coordinate system to ingoing EF one, bulk Yang-Mills field changes as
\begin{align}
	&\tilde C_t(r,t,\vec x) = C_v(r,v,\vec x), \qquad \tilde C_i(r,t,\vec x) = C_i(r,v,\vec x), \nonumber \\
	&\tilde C_r(r,t,\vec x) = C_r(r,v,\vec x) + \frac{C_v(r,v,\vec x)}{r^2f(r)},
\end{align}
where tilded quantities correspond to those in Schwarzschild coordinate system.
The gauge choice \eqref{radial_gauge_Schw} taken throughout this work amounts to imposing $\tilde C_r =0$ in Schwarzschild coordinate system. Then, constraint component of bulk Yang-Mills equation is
\begin{align}
	\tilde \nabla_M \tilde F^{{\rm a}Mr} + \epsilon^{\rm abc} \tilde C_M^{\rm b} \tilde F^{{\rm c} Mr}=0 \Leftrightarrow \nabla_M F^{{\rm a}Mr} + \epsilon^{\rm abc} C_M^{\rm b} F^{{\rm c} Mr}=0
\end{align}
while dynamical EOMs are
\begin{align}
	&\tilde \nabla_M \tilde F^{{\rm a}Mt} + \epsilon^{\rm abc} \tilde C_M^{\rm b} \tilde F^{{\rm c} Mt}=0
	\Leftrightarrow \nabla_M F^{{\rm a}Mv} + \epsilon^{\rm abc} C_M^{\rm b} F^{{\rm c} Mv} \nonumber \\
	& \qquad \qquad \qquad \qquad \qquad \qquad \qquad  - \frac{1}{r^2f(r)} \left[ \nabla_M F^{{\rm a}Mr} + \epsilon^{\rm abc} C_M^{\rm b} F^{{\rm c} Mr} \right] =0, \nonumber \\
	&\tilde \nabla_M \tilde F^{{\rm a}Mi} + \epsilon^{\rm abc} \tilde C_M^{\rm b} \tilde F^{{\rm c} Mi}=0 \Leftrightarrow \nabla_M F^{{\rm a}Mi} + \epsilon^{\rm abc} C_M^{\rm b} F^{{\rm c} Mi}=0. \label{dynamical_EOM_Schw_EF}
\end{align}
Explicitly, the dynamical EOMs \eqref{dynamical_EOM_Schw_EF} in Schwarzschild coordinate system are
\begin{align}
	0 = & \partial_r \left(r^3 \partial_r \tilde C_t^{\rm a } \right) + \frac{1}{r f(r)} \partial_k \left(\partial_k \tilde C_t^{\rm a  } - \partial_t \tilde C_k^{\rm a  } + \epsilon^{\rm abc} \tilde C_k^{\rm b } \tilde C_t^{\rm c } \right) +  \frac{1}{r f(r)} \epsilon^{\rm abc} \tilde C_k^{\rm b } \nonumber\\
	& \times \left(\partial_k \tilde C_t^{\rm c }  - \partial_t \tilde C_k^{\rm c } + \epsilon^{\rm cde} \tilde C_{k}^{\rm d} \tilde C_{t}^{\rm e} \right), \nonumber \\
	0 = & \partial_r\left[ r^3 f(r) \partial_r \tilde C_i^{\rm a } \right] - \frac{1}{r f(r)} \partial_t \left(\partial_t \tilde C_i^{\rm a  } - \partial_i \tilde C_t^{\rm a  } + \epsilon^{\rm abc} \tilde C_t^{\rm b } \tilde C_i^{\rm c } \right)  -\frac{1}{r f(r)} \epsilon^{\rm abc}  \tilde C_t^{\rm b } \nonumber \\
	&\times \left(\partial_t \tilde C_i^{\rm c } - \partial_i \tilde C_t^{\rm c }+ \epsilon^{\rm cde} \tilde C_{t}^{\rm d} \tilde C_{i}^{\rm e} \right) + \frac{1}{r} \partial_k \left(\partial_k \tilde C_i^{\rm a  } -\partial_i \tilde C_k^{\rm a  } + \epsilon^{\rm abc} \tilde C_k^{\rm b } \tilde C_i^{\rm c } \right) \nonumber \\
	&  +  \frac{1}{r} \epsilon^{\rm abc} \tilde C_k^{\rm b } \left(\partial_k \tilde C_i^{\rm c } - \partial_i \tilde C_k^{\rm c }+ \epsilon^{\rm cde} \tilde C_{k}^{\rm d} \tilde C_{i}^{\rm e} \right). \label{dynamical_EOM_Schw_explicit}
\end{align}
In ingoing EF coordinate system, \eqref{dynamical_EOM_Schw_EF} take the following form
\begin{align}
	0 = &  \partial_r \left[ r^3 \left(\partial_v C_{r}^{\rm a  } - \partial_r C_{v}^{\rm a } \right)\right]  -\frac{r}{f(r)} \partial_v \left( \partial_r C_{v}^{\rm a  } - \partial_v C_{r}^{\rm a  }\right) - \frac{1}{rf(r)} \nonumber \\
	& \times \partial_k \left( \partial_k C_{v}^{\rm a  } - \partial_v C_{k}^{\rm a  }+\epsilon^{\rm abc}C_{k}^{\rm b }C_{v}^{\rm c  } \right) - \frac{\epsilon^{\rm abc}}{r f(r)} C_{k}^{\rm b  }  \left(\partial_k C_{v}^{\rm c  } - \partial_v C_{k}^{\rm c  } + \epsilon^{\rm cde} C_{k}^{\rm d} C_{v}^{\rm e} \right) , \nonumber \\
	0 = &\partial_{r }\left[ r^3 f(r) \left(\partial_r C_{i}^{\rm a  } - \partial_{i} C_{r}^{\rm a  } \right)\right] + \partial_r\left[ r \left(\partial_v C_{i}^{\rm a  } - \partial_{i} C_{v}^{\rm a } \right) \right]   \nonumber \\
	&+ r\partial_v \left( \partial_r C_{i}^{\rm a } - \partial_{i} C_{r}^{\rm a } + \epsilon^{\rm abc} C_{r}^{\rm b } C_{i}^{\rm c } \right) + \frac{1}{r} \partial_j \left(\partial_j C_{i}^{\rm a } - \partial_{i} C_{j}^{\rm a } + \epsilon^{\rm abc} C_{j}^{\rm b  } C_{i}^{\rm c } \right) \nonumber \\
	& + r \epsilon^{\rm abc} C_{r}^{\rm b } \left(\partial_v C_{i}^{\rm c  } - \partial_{i} C_{v}^{\rm c  } + \epsilon^{\rm cde} C_{v}^{\rm d} C_{i}^{\rm e} \right)+ \frac{ 1 }{ r } \epsilon^{\rm abc} C_{j}^{\rm b } \left(\partial_j C_{i}^{\rm c  } - \partial_{i} C_{j}^{\rm c  } + \epsilon^{\rm cde} C_{j}^{\rm d} C_{i}^{\rm e} \right). \label{dynamical_EOM_EF_explicit}
\end{align}

When $C_M^{\rm a}$ is linearized as in subsection \ref{YM_linearize}, the dynamical EOMs \eqref{dynamical_EOM_EF_explicit} turn into a system of linear PDEs:
\begin{align}
	&\partial_r\left( r^3\partial_r C_v^{{\rm a}(n)} \right) - \partial_r \left(r^3 \partial_v C_r^{{\rm a}(n)} \right) + \frac{r}{f(r)} \partial_r \partial_v C_v^{{\rm a}(n)} - \frac{r}{f(r)} \partial_v^2 C_r^{{\rm a}(n)} \nonumber \\
	&+ \frac{1}{r f(r)} \left( {\vec\partial}^{\,2} C_v^{{\rm a}(n)} - \partial_v \partial_k C_k^{{\rm a}(n)} \right) = \mathbb S_v^{{\rm a}(n)}, \nonumber \\
	&\partial_r\left[ r^3 f(r)\partial_r C_i^{{\rm a}(n)}\right] - \partial_r \left[r^3 f(r) \partial_i  C_r^{{\rm a}(n)} + r \left( \partial_v  C_i^{{\rm a}(n)} -\partial_i  C_v^{{\rm a}(n)} \right) \right] + r \partial_r \partial_v C_i^{{\rm a}(n)} \nonumber \\
	&-r\partial_v \partial_i C_r^{{\rm a}(n)} + \frac{1}{r} \partial_k \left(\partial_k C_i^{{\rm a}(n)} - \partial_i C_k^{{\rm a}(n)} \right)  = \mathbb S_i^{{\rm a}(n)}, \label{Cmu_perturb_eom_EF}
\end{align}
where the source terms $\mathbb S_\mu^{{\rm a}(n)}$ are products of lower order solutions (as well as their derivatives). For the first two orders, the source terms are
\begin{align}
	\mathbb S_v^{{\rm a}(1)}=& \mathbb S_i^{{\rm a}(1)}=0, \nonumber \\
	\mathbb S_v^{{\rm a}(2)}=& -\frac{ 1 }{ rf(r) }\partial_k \left(\epsilon^{\rm abc} C_k^{\rm b (1)}C_{v}^{\rm c (1)} \right)- \frac{\epsilon^{\rm abc}}{rf(r)} C_k^{\rm b (1)} \left( \partial_k C_{v}^{\rm c (1)} - \partial_v C_k^{\rm c (1)} \right) \nonumber \\
	\mathbb S_i^{{\rm a}(2)}=& -r\partial_v \left( \epsilon^{\rm abc} C_{r}^{\rm b (1)} C_{i}^{\rm c (1)} \right)-\frac{ 1 }{  r} \partial_k \left( \epsilon^{\rm abc} C_k^{\rm b  (1)} C_{i}^{\rm c  (1)} \right)\nonumber \\
	& - r \epsilon^{\rm abc} C_{r}^{\rm b (1) } \left(\partial_v C_{i}^{\rm c (1) } - \partial_{i} C_{v}^{\rm c (1) } \right)- \frac{ 1 }{ r } \epsilon^{\rm abc} C_k^{\rm b (1) } \left(\partial_k C_{i}^{\rm c  (1)} - \partial_{i} C_k^{\rm c (1) } \right). \label{Source_perturb_EF}
\end{align}

In parallel, \eqref{dynamical_EOM_Schw_explicit} are linearized as
\begin{align}
	&\partial_r\left[ r^3\partial_r \tilde C_t^{{\rm a}(n)} \right] + \frac{1}{rf(r)} \partial_k \left(\partial_k \tilde C_t^{{\rm a}(n)} - \partial_t \tilde C_k^{{\rm a}(n)} \right) = \tilde{\mathbb S}_t^{{\rm a}(n)}, \nonumber \\
	&\partial_r\left[ r^3 f(r)\partial_r \tilde C_i^{{\rm a}(n)}\right] - \frac{1}{rf(r)} \partial_t \left(\partial_t \tilde C_i^{{\rm a}(n)} - \partial_i \tilde C_t^{{\rm a}(n)} \right) + \frac{1}{r} \partial_k \left(\partial_k \tilde C_i^{{\rm a}(n)} - \partial_i \tilde C_k^{{\rm a}(n)} \right)= \tilde{ \mathbb S}_i^{{\rm a}(n)},
\end{align}
where for the first two orders the source terms are
\begin{align}
	\tilde{\mathbb S}_t^{{\rm a}(1)}=& \tilde{\mathbb S}_i^{{\rm a}(1)}=0, \nonumber \\
	\tilde{\mathbb S}_t^{{\rm a}(2)}=& - \frac{1}{rf(r)} \partial_k \left(\epsilon^{\rm abc} \tilde C_k^{\rm b (1)} \tilde C_t^{\rm c (1)} \right) -  \frac{1}{rf(r)}\epsilon^{\rm abc} \tilde C_k^{\rm b (1)} \left(\partial_k \tilde C_t^{\rm c (1)}  - \partial_t \tilde C_k^{\rm c (1)} \right), \nonumber \\
	\tilde{\mathbb S}_i^{{\rm a}(2)}=& \frac{1}{rf(r)} \partial_t \left( \epsilon^{\rm abc} \tilde C_t^{\rm b (1)} \tilde C_i^{\rm c (1)} \right) + \frac{1}{rf(r)} \epsilon^{\rm abc}  \tilde C_t^{\rm b (1)} \left(\partial_t \tilde C_i^{\rm c (1)} - \partial_i \tilde C_t^{\rm c (1)}\right) \nonumber \\
	& - \frac{1}{r} \partial_k  \left( \epsilon^{\rm abc} \tilde C_k^{\rm b (1)} \tilde C_i^{\rm c (1)} \right) -  \frac{1}{r} \epsilon^{\rm abc} \tilde C_k^{\rm b (1)} \left(\partial_k \tilde C_i^{\rm c (1)} - \partial_i \tilde C_k^{\rm c (1)} \right) .
\end{align}

\section{More on partially on-shell solution: beyond leading order} \label{generic_structure_solution}

In this appendix we elaborate on generic structure of partially on-shell solution beyond leading order. Recall that the dynamical EOMs at each order in perturbative expansion differ by source terms. Therefore, we will take the next-to-leading order correction $C_\mu^{{\rm a}(2)}$ as the example, for which the source terms do not vanish.

We start with the transverse mode $C_\perp^{{\rm a}(2)}$, which satisfies a closed ODE
\begin{align}
\partial_r\left[r^3f(r)\partial_r C_\perp^{{\rm a}(2)} \right] + \Box_\perp(\partial_r; \omega,q^2)C_\perp^{{\rm a}(2)} = \mathbb S_\perp^{(2)}(r,k^\mu), \label{Cperp_2nd_eom}
\end{align}
where the operator $\Box_\perp(\partial_r; \omega,q)$ and the source term can be read off from \eqref{Cmu_perturb_eom_EF} and \eqref{Source_perturb_EF} in appendix \ref{YM_eom_explicit}. Via Green's function method, \eqref{Cperp_2nd_eom} is solved by
\begin{align}
C_\perp^{{\rm a}(2)}(r,k^\mu) = \int_{\infty_2}^{\infty_1} G_\perp(r,r^\prime;k^\mu) \mathbb S_\perp^{(2)}(r^\prime, k^\mu) dr^\prime, \qquad r\in(\infty_2, \infty_1), \label{Cperp_2nd_solution}
\end{align}
where $G_\perp(r,r^\prime;k^\mu)$ is the Green's function satisfying
\begin{align}
\partial_r\left[r^3f(r)\partial_r G_\perp(r,r^\prime;k^\mu) \right] + \Box_\perp(\partial_r; \omega,q^2) G_\perp(r,r^\prime;k^\mu) = \delta(r-r^\prime).
\end{align}
Thus, $G_\perp$ is indeed the bulk-to-bulk propagator. In order to uniquely fix $G_\perp(r,r^\prime;k^\mu)$, we impose two boundary conditions
\begin{align}
G_\perp(r=\infty_1,r^\prime;k^\mu)=0, \qquad \qquad G_\perp(r=\infty_2,r^\prime;k^\mu)=0,
\end{align}
which is convenient since $C_\perp^{{\rm a}(2)}$ also vanishes at both AdS boundaries.

The Green's function $G_\perp(r,r^\prime;k^\mu)$ could be constructed from linearly independent solutions in \eqref{Cperp_solution_LO}. For convenience, we make linear combination over the two linearly independent solutions presented in \eqref{Cperp_solution_LO} and generate two new basis solutions
\begin{align}
&Y_1(r,k^\mu) = a C_\perp^{\rm ig}(r,k^\mu) + C_\perp^{\rm ig}(r,\bar k^\mu)e^{i\omega \zeta(r)}, \nonumber \\
&Y_2(r,k^\mu)= C_\perp^{\rm ig}(r,k^\mu) + b C_\perp^{\rm ig}(r,\bar k^\mu)e^{i\omega \zeta(r)}, \label{Y_perp}
\end{align}
where we have analytically continued the linearly independent solutions of \eqref{Cperp_solution_LO} so that they are valid over the entire contour. Accordingly, the function $\zeta(r)$ is
\begin{align}
\zeta(r) = \int_{\infty_2}^r \frac{dy}{y^2f(y)}, \qquad r\in(\infty_2, \infty_1),
\end{align}
which is multi-valued. The coefficients $a,b$ in \eqref{Y_perp} are fixed by imposing
\begin{align}
Y_1 (r=\infty_1) = 0, \qquad \qquad Y_2 (r=\infty_2) = 0,
\end{align}
which becomes possible since $\zeta(r=\infty_1) \neq \zeta(r=\infty_2)$. Eventually, the Green's function $G_\perp$ is
\begin{align}
G_\perp(r,r^\prime;k^\mu) = \frac{1}{r^{\prime3} f(r^\prime) W(r^\prime)} \left[ Y_2(r)Y_1(r^\prime) \theta(r^\prime- r) + Y_1(r) Y_2(r^\prime) \theta(r^\prime- r) \right],
\end{align}
where $W(r)$ is the Wronskian determinant of $Y_1(r), Y_2(r)$
\begin{align}
W(r) \equiv Y_2(r) \partial_r Y_1(r) - Y_1(r) \partial_r Y_2(r) = \frac{2i\pi}{r^3f(r)}
\end{align}
The step function $\theta(r-r^\prime)$ is defined on the radial contour of Figure \ref{holographic_SK_contour}, with $r > r^\prime$ ($r < r^\prime$) understood as counter clockwise path-ordered relations.

We turn to the longitudinal sector, which involves a system of two modes $C_v^{{\rm a}(2)}$ and $C_x^{{\rm a}(2)}$. Thus, the above method based on Green's function should be extended appropriately to that based on Green's matrix. We advance by rewriting second order ODEs \eqref{longit_eom} for $C_v^{{\rm a}(2)}$ and $C_x^{{\rm a}(2)}$ into a system of first order ODEs:
\begin{align}
\partial_r X(r,k^\mu) = M(r,k^\mu) X(r,k^\mu) + g(r,k^\mu) \label{Cparallel_eom_1st}
\end{align}
where
\begin{align}
X = \left(C_v^{{\rm a}(2)}, ~ \partial_rC_v^{{\rm a}(2)}, ~C_x^{{\rm a}(2)}, ~\partial_rC_x^{{\rm a}(2)}\right)^{\rm T}
\end{align}
The $4\times 4$ matrix $M$ and column vector $g$ can be directly read off from second order ODEs \eqref{longit_eom}. For simplicity, we will not report their expressions here. First, we consider the homogeneous part of \eqref{Cparallel_eom_1st}
\begin{align}
\partial_r X(r,k^\mu) = M(r,k^\mu) X(r,k^\mu) \label{Cparallel_eom_1st_homo}
\end{align}
The four linearly independent solutions for \eqref{Cparallel_eom_1st_homo} are presented in \eqref{Cparallel_solution_LO}, which we rewrite here
\begin{align}
&X_1 = \left(C_v^{\rm ig}(r,k^\mu), ~\partial_rC_v^{\rm ig}(r,k^\mu), ~C_x^{\rm ig} (r,k^\mu), ~\partial_r C_x^{\rm ig}(r,k^\mu) \right)^{\rm T}, \nonumber \\
&X_2 = \left(C_v^{\rm ig}(r,\bar k^\mu)e^{i\omega\zeta(r)}, ~\partial_r \left( C_v^{\rm ig}(r,\bar k^\mu)e^{i\omega\zeta(r)} \right), ~C_x^{\rm ig}(r,\bar k^\mu) e^{i\omega\zeta(r)}, ~\partial_r \left( C_x^{\rm ig}(r,\bar k^\mu) e^{i\omega\zeta(r)} \right) \right)^{\rm T}, \nonumber \\
&X_3 = \left(C_v^{\rm pg}(r,k^\mu), ~\partial_rC_v^{\rm pg}(r,k^\mu), ~C_x^{\rm pg} (r,k^\mu), ~\partial_r C_x^{\rm pg}(r,k^\mu) \right)^{\rm T}, \nonumber \\
&X_4 = \left(C_v^{\rm pn}(r,k^\mu), ~\partial_rC_v^{\rm pn}(r,k^\mu), ~C_x^{\rm pn} (r,k^\mu), ~\partial_r C_x^{\rm pn}(r,k^\mu) \right)^{\rm T}, \label{X_independent_solution}
\end{align}
where as for the transverse sector we have analytically continued all the linearly independent solutions to the entire radial contour. The linearly independent solutions \eqref{X_independent_solution} for homogeneous system \eqref{Cparallel_eom_1st_homo} help to build a fundamental matrix
\begin{align}
M_0(r,k^\mu) = \left( X_1,~ X_2,~ X_3, ~ X_4 \right)
\end{align}
which satisfies
\begin{align}
\partial_r M_0(r,k^\mu) = M(r,k^\mu) M_0(r,k^\mu).
\end{align}
Then, the general solution for the inhomogeneous system \eqref{Cparallel_eom_1st} is
\begin{align}
X(r,k^\mu) = M_0(r,k^mu) c + M_0(r,k^\mu) \int_{\infty_2}^r M_0^{-1}(r^\prime,k^\mu) g(r^\prime, k^\mu) dr^\prime,
\end{align}
where the column vector $c$ represents the four integration constants, to be determined by boundary conditions. Here, a subtlety arises from boundary conditions for $C_v^{{\rm a}(2)}$: besides vanishing conditions at the AdS boundaries, it is also imposed to be zero at the horizon. Thus, the final solution for longitudinal sector will be piecewise as for the leading order case.

\section*{Acknowledgements}

We would like to thank Gao-Liang Zhou and Tianchun Zhou for helpful discussions.

\bibliographystyle{utphys}
\bibliography{reference}
\end{document}